\documentclass[twocolumn,twocolappendix]{aastex631}

\usepackage{bm}

\usepackage{siunitx}
\usepackage{amsmath}
\usepackage{mathtools}

\usepackage{multirow}
\usepackage{enumitem}

\begin{document}

\title{Kinematics of Supernova Remnants Using Multiepoch Maximum Likelihood Estimation:\\ Chandra Observation of Cassiopeia A as an Example}

\author[0000-0002-5809-3516]{Yusuke Sakai}
\affiliation{Department of Physics, Rikkyo University, Toshima-Ku, Tokyo, 171-8501, Japan}

\author[0000-0003-4808-893X]{Shinya Yamada}
\affiliation{Department of Physics, Rikkyo University, Toshima-Ku, Tokyo, 171-8501, Japan}

\author[0000-0001-9267-1693]{Toshiki Sato}
\affiliation{Department of Physics, School of Science and Technology, Meiji University, 1-1-1 Higashi Mita, Tama-ku, Kawasaki, Kanagawa 214-8571, Japan}

\author[0000-0002-3752-0048]{Ryota Hayakawa}
\affiliation{International Center for Quantum-field Measurement Systems for Studies of the Universe and Particles (QUP), KEK, 1-1 Oho, Tsukuba, Ibaraki 305-0801, Japan}

\author[0000-0001-8335-1057]{Nao Kominato}
\affiliation{Department of Physics, Rikkyo University, Toshima-Ku, Tokyo, 171-8501, Japan}

\begin{abstract}
Decadal changes in a nearby supernova remnant (SNR) were analyzed using a multiepoch maximum likelihood estimation (MLE) approach. To achieve greater accuracy in capturing the dynamics of SNRs, kinematic features and point-spread function effects were integrated into the MLE framework. Using Cassiopeia A as a representative example, data obtained by the Chandra X-ray Observatory in 2000, 2009, and 2019 were utilized. The proposed multiepoch MLE was qualitatively and quantitatively demonstrated to provide accurate estimates of various motions, including shock waves and faint features, across all regions. To investigate asymmetric structures, such as singular components that deviate from the direction of expansion, the MLE method was extended to combine multiple computational domains and classify kinematic properties using the $k$-means algorithm. This approach allowed for the mapping of different physical states onto the image, and one classified component was suggested to interact with circumstellar material by comparison with infrared observations from the James Webb Space Telescope. Thus, this technique will help quantify the dynamics of SNRs and discover their unique evolution.
\end{abstract}

\section{Introduction}
The explosion mechanism of core-collapse supernovae (CCSNe) remains a long-standing problem in astrophysics.
The neutrino-driven explosion is widely accepted as the primary mechanism for CCSNe (e.g., simulations: \citet{janka2016physics,burrows2021core}, observations: \citet{sato2021high}). 
In this mechanism, the explosion is initiated by a core bounce shock that initially stalls but is subsequently revived through neutrino heating. 
Asymmetries, such as hydrodynamic instabilities, are thought to play a critical role in this revival process. A comprehensive understanding of these asymmetric structures necessitates tracking explosive ejections throughout the entirety of supernova remnants (SNRs).

SNRs serve as important laboratories for high-energy phenomena, such as cosmic ray acceleration, magnetic field evolution, the expansion timescales of forward and reverse shocks, and interactions with circumstellar and interstellar material (CSM/ISM). The Chandra X-ray Observatory \citep[hereafter, Chandra;][]{weisskopf2003overview} has achieved a spatial resolution of $\approx$0$''$.5, the highest among existing X-ray telescopes. 
Chandra's combination of superior observational capabilities and long-term operations from 1999 to the present has significantly advanced these fields \citep[e.g.,][]{bamba2005spatial,uchiyama2007extremely,patnaude2009proper,Sato_2018,sato2021high,Matsuda_2022,Tsuchioka_2022}.
Despite these advancements, there remains considerable potential for further leveraging the extensive data accumulated over the decades.

Kinematic measurements can be conducted in two directions: line-of-sight motion via Doppler shift and perpendicular line-of-sight motion (proper motion) by correlating observed images. This paper addresses the following improvements in proper motion measurement:
\begin{enumerate}[noitemsep]
\item Correcting the response function of the observation equipment to maximize the utility of the data (Section~\ref{Image Deconvolution}).
\item Implementing statistical processing and frameworks to fully utilize all available data (Section~\ref{Proper Motion Estimation}).
\item Employing objective data interpretation methods to ensure consistency and reproducibility (Sections~\ref{Result of the MLE Method} and \ref{Magnitude and Standard Deviation of Velocity}).
\item Utilizing classification methods and multi-wavelength data to explore unknown patterns and structures (Sections~\ref{k-means Clustering}, \ref{Discussion}, and \ref{Further Improvement Approach}).
\end{enumerate}
These points are detailed below.

First, Chandra observations are affected by variations in optical axes and roll angles between observations, as well as image degradation due to off-axis aberrations. Correcting these degradations is crucial to maximize the utility of the accumulated data. The image degradation during observation is represented by the point-spread function (PSF). Image deconvolution techniques that use the PSF to estimate the true sky image can help mitigate these effects \citep[review on deconvolution in astronomy by][]{starck2002deconvolution}. In X-ray astronomy, Richardson-Lucy \citep[RL;][]{richardson1972bayesian,lucy1974iterative} deconvolution is frequently employed \citep[e.g.,][]{Grefenstette_2015,Thimmappa_2020,Sobolenko2022,morii2024deconv}. This method typically applies a single PSF across the entire image. However, since the PSF shape varies within the field of view, it is advantageous to account for this effect in diffuse objects such as SNRs. Consequently, we employ an RL with a spatially variant PSF \citep[RL$_{\mathrm{sv}}$;][]{tajima2007studies,tai2010richardson,Sakai_2023} optimized for Chandra \citep{Sakai_2023} as the image preprocessing step for proper motion estimation.

Second, maximum likelihood estimation \citep[MLE;][]{fisher1922mathematical} has been widely utilized in studies measuring the proper motion of SNRs \citep[e.g.,][]{sato2017freely,Sato_2018,millard2020ejecta,Tsuchioka_2021,Tsuchioka_2022}. 
This method is usually applied between two observations, but it can also be employed in a framework involving multiple observations. 
For instance, \citet{vink2022forward} implemented a multiepoch MLE optimized for forward and reverse shock measurements that incorporates expansion constraints. In this study, we implement a multiepoch MLE with appropriate motion constraints applicable across the full range of SNRs. Although several methods exist for analyzing the dynamics of SNRs, including optical flow \citep{farneback2003two} as used by \citet{Sato_2018} and B-spline techniques \citep{lee1996image,lee1997scattered} as utilized by \citet{ichinohe2023spatiotemporal}, the MLE method can be easily optimized and scaled to meet the specific requirements.

For the last two points, we estimate the proper motion and its uncertainty for each pixel in the entire SNR to ensure data consistency and reproducibility of the MLE results. 
Data from the James Webb Space Telescope \citep[JWST;][]{Rigby_2023} can be used to compare with our findings and gain a better understanding of the interaction with the CSM/ISM. Furthermore, unsupervised classification methods can offer insights into the complex kinematics of SNRs from an objective standpoint. While numerous previous studies have identified salient features \citep[e.g.,][]{uchiyama2007extremely,uchiyama2008fast,patnaude2006small,patnaude2014comparison,matsuda2020temporal,Matsuda_2022}, our approach aims to reveal the dynamics across the entire range of them in a reproducible and data-driven manner.

The structure of this paper is as follows. 
Section~\ref{Method} describes the formula for RL$_{\mathrm{sv}}$ deconvolution and multiepoch MLE.
We constrain the motion direction and magnitude around these images to estimate various types of motion using our multi-image MLE. In Section~\ref{Application to Observed Data}, we apply the MLE to the Galactic SNR Cassiopeia A (Cas~A) using data obtained by Chandra in 2000, 2009, and 2019. Section~\ref{Discussion} compares our proper motion results with previous studies, including forward and reverse shock velocities and expansion rates. We also objectively examine the movement of features away from the explosion center and changes in brightness. Furthermore, we present enhancements to the proposed MLE, including methods to more accurately capture both local and global motion and to classify physical states based on the obtained motion. Finally, Section~\ref{Conclusion} concludes the paper.

\begin{figure*}[ht!]
 \centering
 \includegraphics[width=1\linewidth]{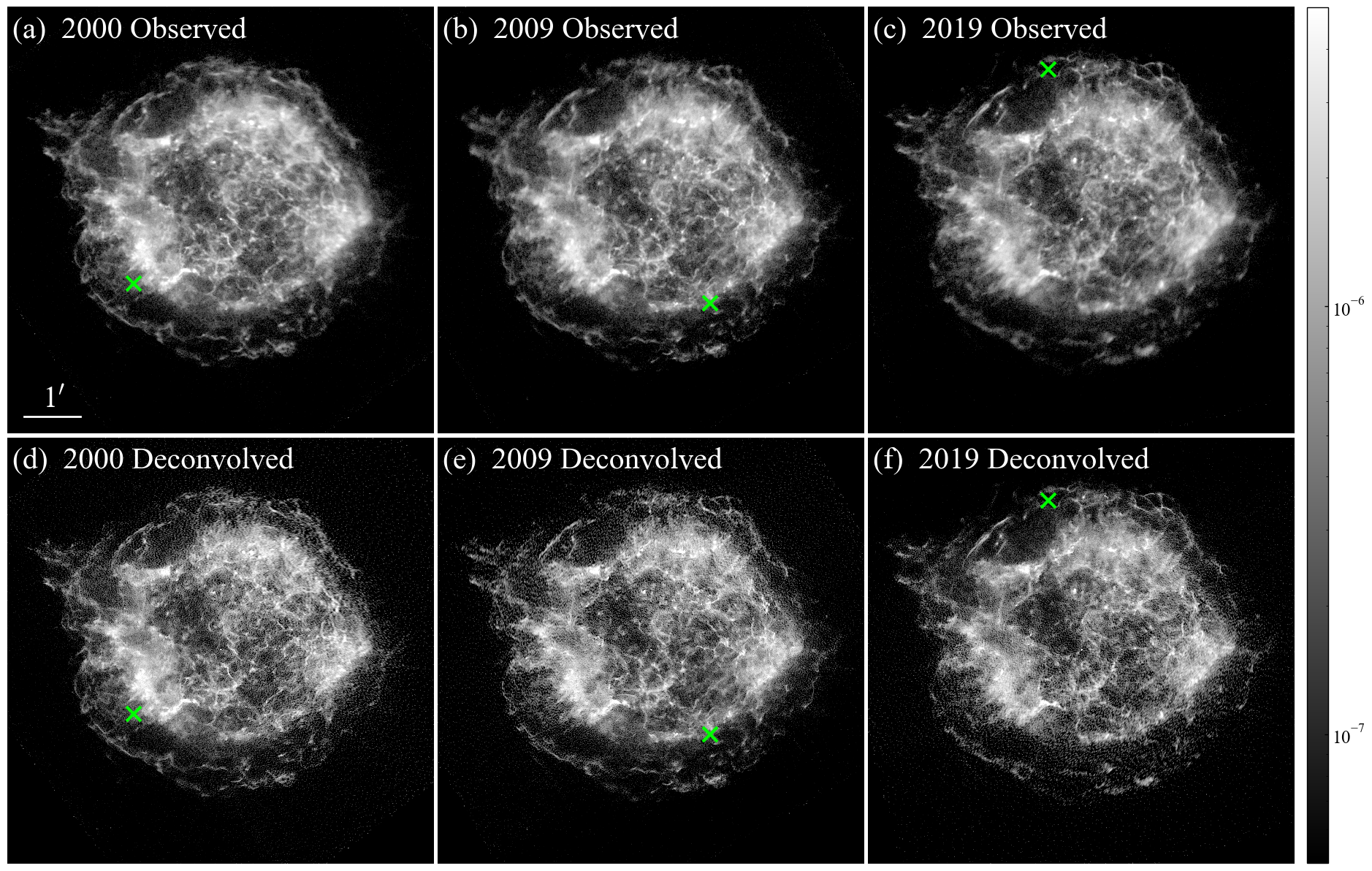}
 \caption{(a, b, c): X-ray observed images in the 2--7 keV band of Cas~A obtained with Chandra in the years 2000, 2009, and 2019, respectively. (d, e, f): Applying the RL$_{\mathrm{sv}}$ deconvolution in each of (a, b, c). The unit of flux in the images is \si{photons.cm^{-2}.pixel^{-2}.s^{-1}}. The optical axis of each panel is indicated by a cross.}
 \label{rlsv}
\end{figure*}

\section{Method}\label{Method}
\subsection{Image Deconvolution}\label{Image Deconvolution}
\subsubsection{RL Deconvolution}
The RL algorithm iteratively estimates the true image from an observed image using Bayesian inference. This method generally assumes a single-shape PSF, meaning the PSF does not change with position in the image. The algorithm is given by
\begin{equation}
W_{i}^{(r+1)} = W_{i}^{(r)} \sum_k \frac{P_{ik}H_k}{\sum_j P_{jk}W_{j}^{(r)}},
\label{rl_eq}
\end{equation}
where $i$ and $j$ map the image in the sky and $k$ maps the image on the detector. The indices of the summation run through all pixels. $W^{(r)}$ is the restored image after $r$ iterations, and $H$ is the observed image at the ACIS detector. $P_{jk}$ is the probability that a photon emitted in sky bin $j$ will be measured in data space bin $k$, or $P(H_k|W_j)$.

\subsubsection{RL with a Spatially Variant PSF}
The RL$_{\mathrm{sv}}$ method is based on the RL algorithm but uses a spatially variant PSF. The method is expressed by
\begin{equation}
W{i}^{(r+1)} = W_{i}^{(r)} \sum_k \frac{P_{iik}H_k}{\sum_j P_{jjk}W_{j}^{(r)}},
\label{rlsv_eq}
\end{equation}
where $P_{jjk}$ refers to the PSF at position $j$ (first index), representing the probability that an event emitted at $W_j$ (second index) will be observed at $H_k$ (third index), or $P_j(H_k|W_j)$. The computational cost and memory requirements need to be minimized for calculating the third-order tensor of the PSF, which is a distinctive feature of the RL$_{\mathrm{sv}}$ algorithm. When $H$ is corrected for slight differences among pixels in effective area and exposures, its normalization can be chosen arbitrarily. Here we use $H_k = N_k / A_k$, where $N_k$ is the detector count image and $A_k$ is the Hadamard product of effective area and exposure time.

\begin{figure*}[ht!]
 \centering
 \includegraphics[width=1\linewidth]{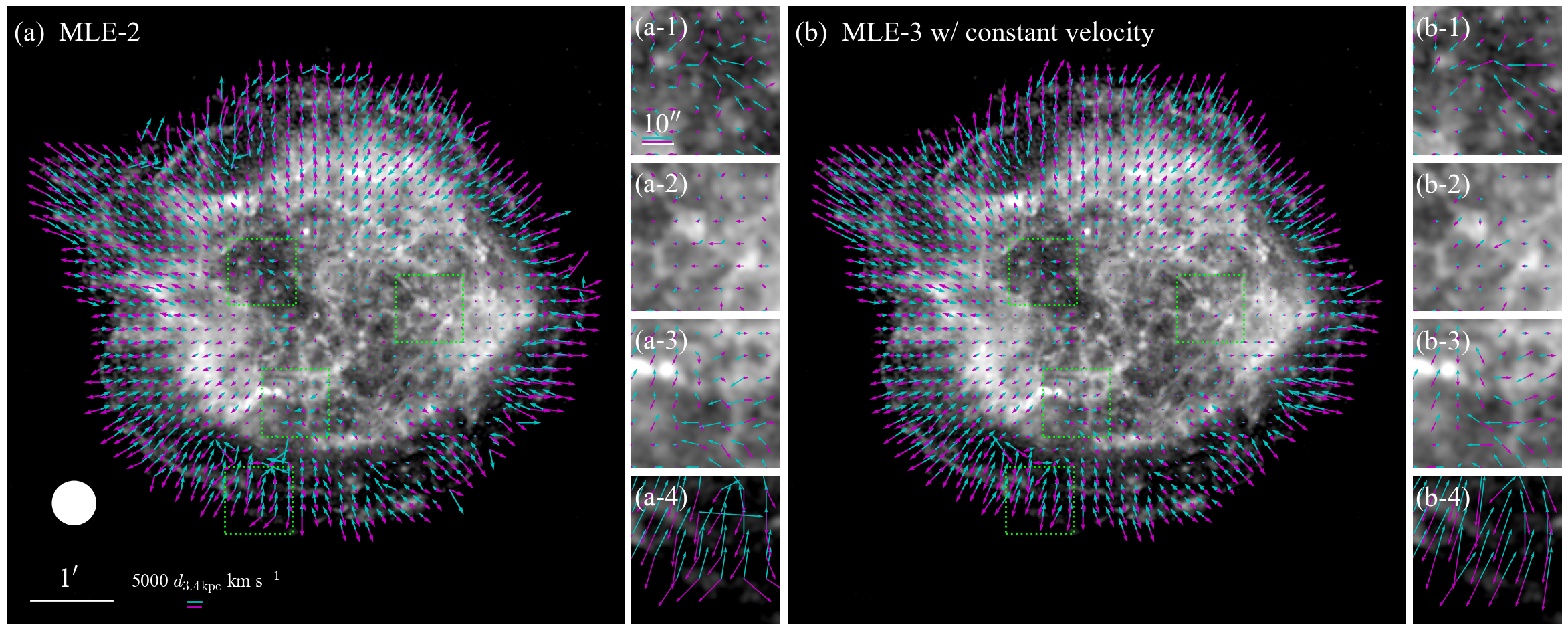}
 \caption{Two-dimensional proper motion results calculated by MLE methods. (a): The result of MLE-2 proper motions with the years 2009 and 2000, and 2009 and 2019, independently. The cyan and magenta vectors represent the proper motion from 2009 to 2000, and from 2009 to 2019, respectively. These vectors are normalized to 30-year units and displayed at 20-pixel intervals. The background image is from 2009. The circular area (65 pixels in diameter) displayed in (a) represents the computation domain used for the MLE calculation.  (a-1, -2, -3, -4): Enlarged images indicated by the colored frames in (a), with vectors aligned at 15-pixel intervals. (b): The same as in (a), except for applying the MLE-3 with constant velocity for 2000, 2009, and 2019.}
 \label{proper_motion}
\end{figure*}

\subsection{Proper Motion Estimation}\label{Proper Motion Estimation}
Proper motion estimation can be approached in two ways: a one-dimensional method that creates a profile with a specified motion direction in advance to obtain statistics \citep[e.g.,][]{DeLaney_2003,katsuda2010x,tanaka2020shock,suzuki2022particle}, and a two-dimensional method that estimates motion from image information \citep[e.g.,][]{sato2017freely,millard2020ejecta,Tsuchioka_2021,Tsuchioka_2022}. The former is not suitable for estimating the dynamics in the entire region of the remnant because it is difficult to preselect the direction. Therefore, in this paper, the latter approach is used.

\subsubsection{MLE for a Poisson Process} \label{MLE for a Poisson Process}
The calculation of two-dimensional proper motion using MLE is derived from the following assumptions. If the changes in the spatial distribution and flux between two count images taken at different times are invariant within the statistical error and parameterized solely by the displacement, they can be modeled by a Poisson function.
Here, two epoch images are denoted as epoch-0 and epoch-1, photon count images are given by $N^{(0)},\,N^{(1)}$, and the Hadamard product of effective area and exposure time maps are represented by $A^{(0)},\,A^{(1)}$. The likelihood function is formulated as
\begin{equation}
\mathcal {L}\left(\bm{\Delta}_{x,y};\,\Lambda,\,K\right) = \prod_{(x,y)\in \Omega} \frac{e^{-\Lambda_{x,y}} \Lambda_{x,y}^{K_{x+\Delta_x,y+\Delta_y}}}{K_{x+\Delta_x,y+\Delta_y}!},
 \label{mle_eq}
\end{equation}
where $\Lambda$ and $K$ represent different time points of observation denoted as $\Lambda=\frac{A^{(0)}}{A^{(1)}}N^{(1)}$ (normalized by same exposure as $N^{(0)}$) and $K=N^{(0)}$.
The $\Lambda$ is fitted to the Poisson statistics as a model, with $K$ serving as actual measurements.
The $\Lambda_{x,y}$ represents the pixel value at the coordinates $(x,\,y)$ in the image of $\Lambda$.
Note that $\Lambda_{x,y}=0$ is undefined in the Poisson distribution, so it is replaced by a smaller value of 0.01.
$(x+\Delta_x,\,y+\Delta_y)$ denotes the coordinates shifted by the vector $\bm{\Delta}_{x,y}= (\Delta_x,\,\Delta_y)$.
The $\Omega$ represents the computation region. The proper motion is obtained by specifying the range of $\bm{\Delta}_{x,y}$ and searching for the value that maximizes $\mathcal {L}(\cdot)$.
In this paper, $(x,\,y)$ represents the pixel coordinates of the ACIS detector (one unit pixel corresponds to 0$''$.492), and the search for $\bm{\Delta}_{x,y}$ is performed in pixel units.
This method is denoted as the MLE of two observations, abbreviated as the MLE-2 method.

\begin{table*}[ht!]
 \caption{Basic Information on the Chandra Observations of Cas~A Used in this Paper}
  \centering
   \begin{tabular}{ccccccccc} \hline\hline
      Obs. ID & Obs. Start & Exp. Time & Detector & R.A. & Decl. & Roll &\multicolumn{2}{c}{MLE Usage}\\
       & (yyyy~mmm~dd) & (ks) & & (deg) & (deg) & (deg) & Uniform Domain & Preselected Domain \\ \hline
       114 & 2000 Jan 30 & 49.93 & ACIS-S & 350.9159 & 58.7926 &  323.3801 & \checkmark &  \\ \hline
       4636 & 2004 Apr 20 & 143.48 & ACIS-S & 350.9129 & 58.8412 & 49.7698& & \checkmark\\
       4637 & 2004 Apr 22 & 163.50 & ACIS-S & 350.9131 & 58.8414 & 49.7665& & \checkmark\\
       4639 & 2004 Apr 25 & 79.05 & ACIS-S & 350.9132 & 58.8415 & 49.7666 & & \checkmark\\
       5319 & 2004 Apr 18 & 42.26 & ACIS-S & 350.9127 & 58.8411 &  49.7698& & \checkmark\\ \hline
       10935 & 2009 Nov 2 & 23.26 & ACIS-S & 350.8329 & 58.7868 &  239.6794 & \checkmark& \\
       12020 & 2009 Nov 3 & 22.38 & ACIS-S & 350.8330 & 58.7871 &  239.6796 & \checkmark& \\ \hline
       19606 & 2019 May 13 & 49.42 & ACIS-S & 350.8854 & 58.8559 &  75.1398 & \checkmark&\checkmark \\ \hline
   \end{tabular}
  \label{used data}
\end{table*}

\subsubsection{MLE with Three Observations}
When computing the proper motion among three consecutive observations, denoted as epoch-0, epoch-1, and epoch-2, and using the center frame of epoch-1 as the reference model, an extended formulation of Equation~\eqref{mle_eq} can be expressed as follows:
\begin{equation}
\begin{split}
&\mathcal{L}_{\textrm{MLE-3}}\left(\bm{\Delta}^{(10)}_{x,y},\,\bm{\Delta}^{(12)}_{x,y};\, \Lambda,\, K^{(0)},\, K^{(2)}\right)\\
&= \mathcal{L}\left(\bm{\Delta}^{(10)}_{x,y};\, \Lambda^{(10)},\, K^{(0)}\right) \mathcal{L}\left(\bm{\Delta}^{(12)}_{x,y};\, \Lambda^{(12)},\, K^{(2)}\right)\\
&\quad\times R\left(\bm{\Delta}^{(10)}_{x,y},\, \bm{\Delta}^{(12)}_{x,y}\right),
 \label{mle_3_eq}
 \end{split}
\end{equation}
where each variable represents $K^{(0)}=N^{(0)}$, $K^{(2)}=N^{(2)}$, $\Lambda^{(10)}=\frac{A^{(0)}}{A^{(1)}}N^{(1)}$, and $\Lambda^{(12)}=\frac{A^{(2)}}{A^{(1)}}N^{(1)}$. 
The displacement from epoch-1 to epoch-0 and from epoch-1 to epoch-2 are denoted as $\bm{\Delta}^{(10)}_{x,y}$ and $\bm{\Delta}^{(12)}_{x,y}$, respectively. The $\mathcal{L}(\cdot)$ is the same as Equation~\eqref{mle_eq}, and $R(\cdot)$ is the regularization term applied to the observation positions. In the absence of $R(\cdot)$, i.e., when $R(\cdot)$ is constant, the maximum likelihood solution of this equation is equivalent to the independently computed solution of Equation~\eqref{mle_eq}.

By incorporating appropriate constraints into the solution search region, higher estimation accuracy is expected compared to the MLE-2 method. For application to the entire SNR, we consider constraints that allow for a variety of motions, such as expansion, contraction, and deviation from expansion. Specifically, we introduce two constraints between the measured proper motions: the constant motion direction assumption and the constant velocity assumption.

The directional regularization assumes that the three points ($\bm{\Delta}^{(10)}_{x,y}$, origin $\bm{0}$, and $\bm{\Delta}^{(12)}_{x,y}$) are in the same direction. To determine whether they are on the same line in the digitized image, the Bresenham's line algorithm \citep{bresenham1998algorithm} is used: coordinates $\bm{\Delta}^{(10)}_{x,y}$ and $\bm{\Delta}^{(12)}_{x,y}$ are connected by a Bresenham line, and the maximum likelihood solution is calculated among those that pass through the origin $\bm{0}$. This approach is denoted as MLE-3 without constant velocity.

This MLE-3 method without assuming constant velocity also provides acceleration information. \citet{vink2022forward} calculated accelerations of a few tens of \si{km.s^{-2}} in forward shocks using subpixelization and optimization for the shock waves. 
Although addressing the combination of the RL$_{\mathrm{sv}}$ method and the subpixelization process is beyond the scope of this paper, our pixel-level estimation approach poses challenges for obtaining accurate acceleration measurements. Therefore, the maximum likelihood solution is obtained with the constraint that the acceleration term is excluded, assuming constant velocity. In this paper, since the observed data are almost equally spaced (see Table~\ref{used data}), $\bm{\Delta}^{(10)}_{x,y}=-\bm{\Delta}^{(12)}_{x,y}$ is used in the pixel-wise estimation as a constraint for constant velocity. We examine the validity of this limitation. Table~\ref{used data} shows that the difference in time intervals 2009 to 2000 and 2009 to 2019 is \SI{\approx 2}{\%}. In contrast, the migration of Cas~A over the 10 years period is $\lessapprox$10 pixels, and a difference of 1 pixel represents an error of $\gtrapprox$10\%. Thus, the error in the time difference is within 1 pixel, and the constraint is reasonable for pixel-level accuracy. This method is referred to as MLE-3 with constant velocity.
Based on this insight, MLE-3 with constant velocity is used for the constraints in this paper (MLE-3 without constant velocity is only used in Section~\ref{Optimal Computational Domain Size Estimation} for comparison of constraint accuracy).

\subsection{K-means Clustering}\label{k-means Clustering}
SNRs have various features such as expansion rate, proper motion, and X-ray brightness. These features can be classified using the $k$-means algorithm \citep{macqueen1967some}. This algorithm is an unsupervised clustering method that categorizes $N$ features into $k$ clusters. It is widely applied in clustering tasks, including in the study of SNRs \citep[e.g.,][]{Sasdelli2016,chatzopoulos2019systematic,ichinohe2023spatiotemporal}. The algorithm initially assigns $k$ centroids randomly in an $N$-dimensional space, which can reduce classification accuracy if they are too close together.

To address this issue, $k$-means++ \citep{arthur2007k} enhances $k$-means by spacing initial centroids based on distance. Hence, we use $k$-means++ for better robustness. Although $k$-means++ only differs in its initialization, we refer to it as $k$-means. The method is implemented in the Python scikit-learn package.\footnote{The scikit-learn code is available at \url{https://scikit-learn.org/stable/}.}

\begin{figure*}[ht!]
 \centering
 \includegraphics[width=1\linewidth]{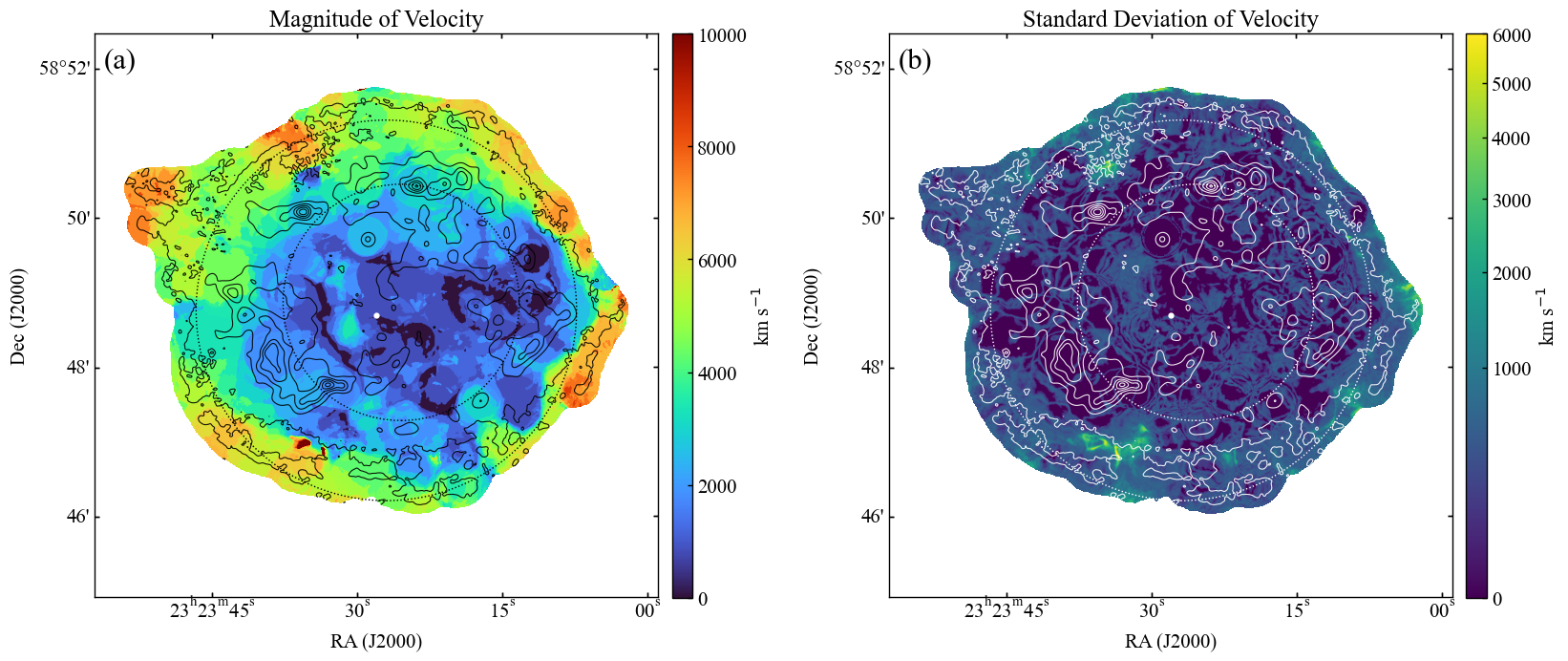}
 \caption{Results of the velocity and variability of proper motion calculated by the MLE-3 method with constant velocity. (a): Magnitude of the velocity field. (b): Standard deviation of the velocity. The contours correspond to the year 2009, and the CCO is marked with a dot. The two dotted line circles represent the positions of the reverse shock and forward shock wave. These results were obtained by applying the MLE-3 method to 100 random images perturbed by statistical errors.}
 \label{statistical_error_map}
\end{figure*}

\section{Application to Observed Data}\label{Application to Observed Data}

\subsection{Data Selection}\label{Data Selection}
We utilize the SNR Cas~A to demonstrate our proposed MLE method. This remnant is one of the best-studied SNRs, with a well-estimated age \citep[$\approx$350 years;][]{thorstensen2001expansion}, distance \citep[$3.4^{+0.3}_{-0.1}$ kpc;][]{reed1995three,alarie2014hyperspectral}, and explosion center \citep[$\alpha$(J2000)=23$^\textrm{h}$23$^\textrm{m}$27$^\textrm{s}$.77$\pm$0$^\textrm{s}$.05, $\delta$(J2000)=58$^\circ$48$'$49$''$.4$\pm$0$''$.4;][]{thorstensen2001expansion}. These characteristics make it an ideal candidate for investigating velocity estimates, expansion rates, and asymmetry structures, which will be discussed in Section~\ref{Discussion}. Moreover, Chandra has observed this object multiple times since its launch in 1999, enabling us to apply MLE over a 20-year period. The rich observations in 2004 \citep{hwang2004million} are particularly valuable for validating our MLE approach.

The Chandra data of Cas A used in this paper are listed in Table~\ref{used data}, comprising ACIS-S observations from 2000 (Obs.~ID=114), 2004 (Obs.~ID=4636, 4637, 4639, 5319), 2009 (Obs.~ID=10935, 12020), and 2019 (Obs.~ID=19606), which are available at \dataset[DOI: 10.25574/cdc.280]{https://doi.org/10.25574/cdc.280}. 
The exposure times are \SI{\approx 50}{ks} for the years 2000, 2009, and 2019, and a significantly longer \SI{\approx 400}{ks} for the year 2004. 
The 2004 image is used to validate the accuracy of the proposed uniform domain MLE-3 by comparing it with the preselected domain MLE-2, as described in Appendix~\ref{Comparison of MLE Computation Domain Uniform and Preselected}. 
The data are organized into nearly uniform intervals, spanning 9.75 years from 2000 to 2009 and 9.54 years from 2009 to 2019.

Data processing and analysis were performed using Chandra Interactive Analysis of Observations \citep[CIAO;][]{fruscione2006ciao} version 4.15 and CALDB version 4.10.4. 
The data were reprocessed from the level 1 event file using \texttt{chandra\_repro}. Aligning Cas~A between ACIS observations presents several challenges. Position and roll angle corrections require at least two sources, but each observation often lacks common sources, making calibration difficult \citep{Holland-Ashford_2024}. 
Although a method using the northeast jet for roll angle estimation has been proposed \citep{DeLaney_2003}, it is hampered by significant changes in spatial distribution over decades. 
The central compact object (CCO) has a proper motion of approximately \SI{\approx300}{km.s^{-1}} \citep{Holland-Ashford_2024}, resulting in variations of less than 1 pixel over 20 years. Although the roll angle cannot be corrected, alignment is often performed using this object \citep[e.g.,][]{Sato_2018,vink2022forward}, and this approach was also employed in this paper. The CCO's position was identified with \texttt{wavdetect}, and the event file was updated with respect to Obs.~ID=114 using \texttt{reproject\_aspect}. The 2004 data were merged with Obs.~ID=4636, 4637, 4639, and 5319 using \texttt{merge\_obs}, since their optical axes are aligned. Similarly, \texttt{merge\_obs} was applied to the 2009 data of Obs.~ID=10935 and 12020. The energy band of 2--7 keV was adopted to mitigate contamination effects at lower energies \citep{plucinsky2018complicated}. This energy band includes both thermal bremsstrahlung and nonthermal (synchrotron) radiation. The images were created using \texttt{flux\_img}.

\subsection{Generating the PSF}
The Chandra telescope system consists of four pairs of nested reflecting surfaces arranged in the Wolter type I geometry. 
The high-energy response is achieved by coating the mirrors with iridium. It holds the record for the highest angular resolution among existing X-ray telescopes, achieving 0$''$.492. Extensive calibration of the Chandra mirror has been conducted both on the ground and in orbit \citep{jerius2000orbital}. The Chandra PSF is position-dependent, mainly due to aberrations.

The Chandra implementation of the RL$_{\mathrm{sv}}$ method by \citet{Sakai_2023}, which utilizes PSFs for each position, samples PSFs at intervals of several tens of pixels. Consequently, we employed PSFs sampled at intervals of 25$\times$25 pixels for each year. For the merged data from 2004 and 2009, Obs.~ID 4636 from 2004 and Obs.ID 10935 from 2009 were used as representative PSFs. The simulation of the PSF utilized the AXAF Response to X-rays model \citep{wise1997simulated,davis2012raytracing}. This model was employed through CIAO's \texttt{simulate\_psf}. The simulations were performed at a monochromatic energy of 3.8 keV, which is the appropriate energy for this band according to the official CIAO page.\footnote{\url{https://cxc.cfa.harvard.edu/ciao/why/monochromatic_energy.html}} The generated PSFs for each location are shown in Appendix~\ref{Position dependence of PSF}.

\subsection{Result of the RL$_{sv}$ Method}\label{Result of the RLsv Method}
Figures~\ref{rlsv}(a, b, c) present the observation images from 2000, 2009, and 2019 in the 2--7 keV band, and Figures~\ref{rlsv}(d, e, f) show the application of the RL$_{\mathrm{sv}}$ method with 50 iterations, respectively. 
The X mark in each panel indicates the optical axis. There is almost no difference between the observed and RL$_{\mathrm{sv}}$ images around the optical axis (on-axis), while the RL$_{\mathrm{sv}}$ image becomes clearer away from the optical axis (off-axis).

Off-axis images in the RL$_{\mathrm{sv}}$ method are less accurate than on-axis images, resulting in reduced spatial resolution. To improve the estimation accuracy of the MLE method, it is essential to match the resolution of the RL$_{\mathrm{sv}}$ images. We therefore convolved the RL$_{\mathrm{sv}}$ results (Figures~\ref{rlsv}(d, e, f)) with a Gaussian distribution with a sigma of 1.5 pixels. The choice of the sigma value is determined based on the off-axis PSF (see Appendix~\ref{Appropriate smoothed Gaussian sigma of RLsv} for more details). These processed images, referred to as smoothed-RL$_{\mathrm{sv}}$ images, are used in the MLE method.

\subsection{Result of the MLE Method}\label{Result of the MLE Method}
We present the results of applying MLEs (MLE-2 and MLE-3 with constant velocity) to the smoothed-RL$_{\mathrm{sv}}$ images. 
In the MLE-2 method, the years $(\Lambda=2009,\,K=2000)$ and $(\Lambda=2009,\,K=2019)$ are used in Equation~\eqref{mle_eq}. 
For the MLE-3 with constant velocity, $(K^{(0)}=2000,\,\Lambda=2009,\,K^{(2)}=2019)$ is substituted in Equation~\eqref{mle_3_eq}.
Since smoothed-RL$_{\mathrm{sv}}$ image is not an integer, $K$ is rounded to an integer. A circular area with a diameter of 65 pixels is used for the computational domain $\Omega$. 
The appropriate calculation region will be discussed in Section~\ref{Optimal Computational Domain Size Estimation}.
The search ranges of $\bm{\Delta}^{(10)}_{x,y}$ and $\bm{\Delta}^{(12)}_{x,y}$ are the same for all regions, with a radius of 15 pixels, which is equivalent to estimating up to $12000$ \si{km.s^{-1}} at a distance of \SI{3.4}{kpc}.
This velocity corresponds to a sufficient coverage of all proper motions in X-rays based on previous studies \citep{vink2022forward}. 

Figures~\ref{proper_motion}(a, b) show the proper motion results of the MLE-2 method and the constant-velocity MLE-3 method, respectively.
Each vector is normalized to 30 years for clarity and displayed every 20 pixels.
The cyan and magenta vectors represent the proper motions for the years 2009 to 2000 and 2009 to 2019, respectively. 
Moving 1 pixel (or 0$''$.492) from 2009--2000 and from 2009--2019 corresponds to \SI{813.7}{km.s^{-1}} and \SI{831.5}{km.s^{-1}}, respectively.
The magnitude of each scaled \SI{5000}{km.s^{-1}} vector is shown in Figures~\ref{proper_motion}(a) and \ref{proper_motion}(a-1). 
We define the outer background region as the average brightness in the region of a 65-pixel circle diameter less than \SI{4e-8}{photons.cm^{-2}.pixel^{-2}.s^{-1}}.
In all images in this paper, this background mask is excluded from the proper motion result.

The MLE-2 (Figure~\ref{proper_motion}(a)) and the constant-velocity MLE-3 (Figure~\ref{proper_motion}(b)) show no significant qualitative differences overall.
For detailed comparison, enlarged images of the colored frames in Figures~\ref{proper_motion}(a, b) are shown in the corresponding panels to the right of the main panel.
The constant-velocity MLE-3 case generally captures all motion, including complex motion \citep{patnaude2009proper} in Figures~\ref{proper_motion}(b-1, -3), inward reverse shocks \citep{Sato_2018} in Figure~\ref{proper_motion}(b-2), and sparse regions in Figure~\ref{proper_motion}(b-4).
A quantitative comparison of these is presented in Section~\ref{Optimal Computational Domain Size Estimation}.

\subsection{Magnitude and Standard Deviation of Velocity}\label{Magnitude and Standard Deviation of Velocity}
We present the velocity magnitude for each pixel and evaluate fluctuations caused by statistical errors in the constant-velocity MLE-3 method.
Figures~\ref{statistical_error_map}(a, b) show the magnitude of the velocity field and its standard deviation from the 2009--2019 side of Figure~\ref{proper_motion}(b). 
The procedure for calculating the standard deviation is described in Appendix~\ref{Procedure for Calculating the Standard Deviation of Velocity}.
The reason for not showing the year 2009--2000 side is that under the addition of the constant velocity constraint, the only difference is in the observed time interval, which provides almost identical velocity information.
The two dotted circles correspond to the positions of the forward shock and reverse shock regions, starting from the outside \citep{gotthelf2001chandra}.

Figure~\ref{statistical_error_map}(a) shows that the velocities of the forward shock regions are approximately 5000--6500 \si{km.s^{-1}}, and the reverse shock regions are about 1500--3500 \si{km.s^{-1}}, which is largely consistent with previous studies \citep[e.g.,][]{delaney2003first,patnaude2009proper,vink2022forward,Wu_2024}. Figure~\ref{statistical_error_map}(b) shows that at the boundaries of the region, estimation is difficult and has an error of about 5 pixels, but elsewhere the determination is generally accurate to within 1 pixel. Detailed validation of these results will be discussed in Section~\ref{Verification of MLE Results with Previous Studies}.

\section{Discussion}\label{Discussion}
\subsection{Verification of MLE Results with Previous Studies}\label{Verification of MLE Results with Previous Studies}
To investigate the stability and applicability of the constant-velocity MLE-3 method, we compared the proper motion velocities in Figure~\ref{statistical_error_map} with those from previous studies. The results are generally consistent with the entire shock wave region of Cas~A measured by \citet{vink2022forward}.

We conducted a comparative analysis of the velocities of the reverse shocks in the C1, C2, W1, W2, W3, and W4 regions as described in \citet{Sato_2018}. The results for the central regions (C1, C2) show consistency. However, in regions W1 and W2 (around Figure~\ref{proper_motion}(b-2)), the velocities are estimated to be lower, ranging from 1500 to 2500 \si{km.s^{-1}}, and in regions W3 and W4 the velocities are estimated to be even lower, between 0 and 1000 \si{km.s^{-1}}. This discrepancy can be attributed to the optimization of the computational domain in \citet{Sato_2018} for motion scales of 21$\times$21 or 31$\times$31 pixels, whereas our study utilizes a larger uniform domain, which smooths out the proper motion and results in lower velocity measurements. 
In Section~\ref{MLE with Appropriate Computational Domains}, we propose a method to address this issue by performing MLE with appropriate computational domains at each location.

\subsection{Comparison with SNR Evolution Model}\label{Comparison with SNR Evolution Model}
The evolution of SNRs is characterized by their expansion timescale. Measuring this provides insights into the asymmetric distribution of ejecta and CSM.
For ejecta-dominated remnants, the expansion parameter $m$ is described by a self-similar evolution with $R \propto t^m$ with $m = (n - 3)/(n - s)$ for a remnant distance $R$, remnant age $t$, power-law index of ejecta density $n$ ($\rho_{\textrm{ej}} \propto r^{-n}$), and power-law index of the CSM density $s$ ($\rho_{\textrm{CSM}}\propto r^{-s}$). Here $s=0$ denotes a uniform density profile, while $s=2$ represents a density profile of the stellar wind. The CCSNe such as Cas~A is expected to have $s=2$.

Cas A is thought to have exploded in 1672 \citep{thorstensen2001expansion}, and such young SNRs are transitioning from the ejecta-dominated phase to the Sedov--Taylor phase \citep{Truelove_1999}.
At this stage, the forward shock position is modeled by $R_\textrm{fs} \propto t^{2/(5-s)}$, where $R_\textrm{fs}$ is the radius of the forward shock wave. Substituting $s=2$ for the stellar wind gives $m=2/3$.
Referring to previous proper motion results for the Cas A expansion in X-rays, $m\approx 0.7$ \citep{delaney2003first,patnaude2009proper,vink2022forward}, which is close to $m = 2/3$.
In reality, however, as shown by the image distribution in Figure~\ref{rlsv} and the proper motions in Figure~\ref{proper_motion}, the densities $\rho_{\textrm{ej}} \propto r^{-n}$ and $\rho_{\textrm{CSM}}\propto r^{-s}$ should be asymmetric, with different expansion rates at different locations.
Therefore, we measured the expansion coefficient from the proper motion result in Figure~\ref{proper_motion}(b) assuming the explosion center at $\alpha$(J2000)=23$^\textrm{h}$23$^\textrm{m}$27$^\textrm{s}$.77, $\delta$(J2000)=58$^\circ$48$'$49$''$.4.

\begin{figure}[ht!]
 \centering
 \includegraphics[width=1\linewidth]{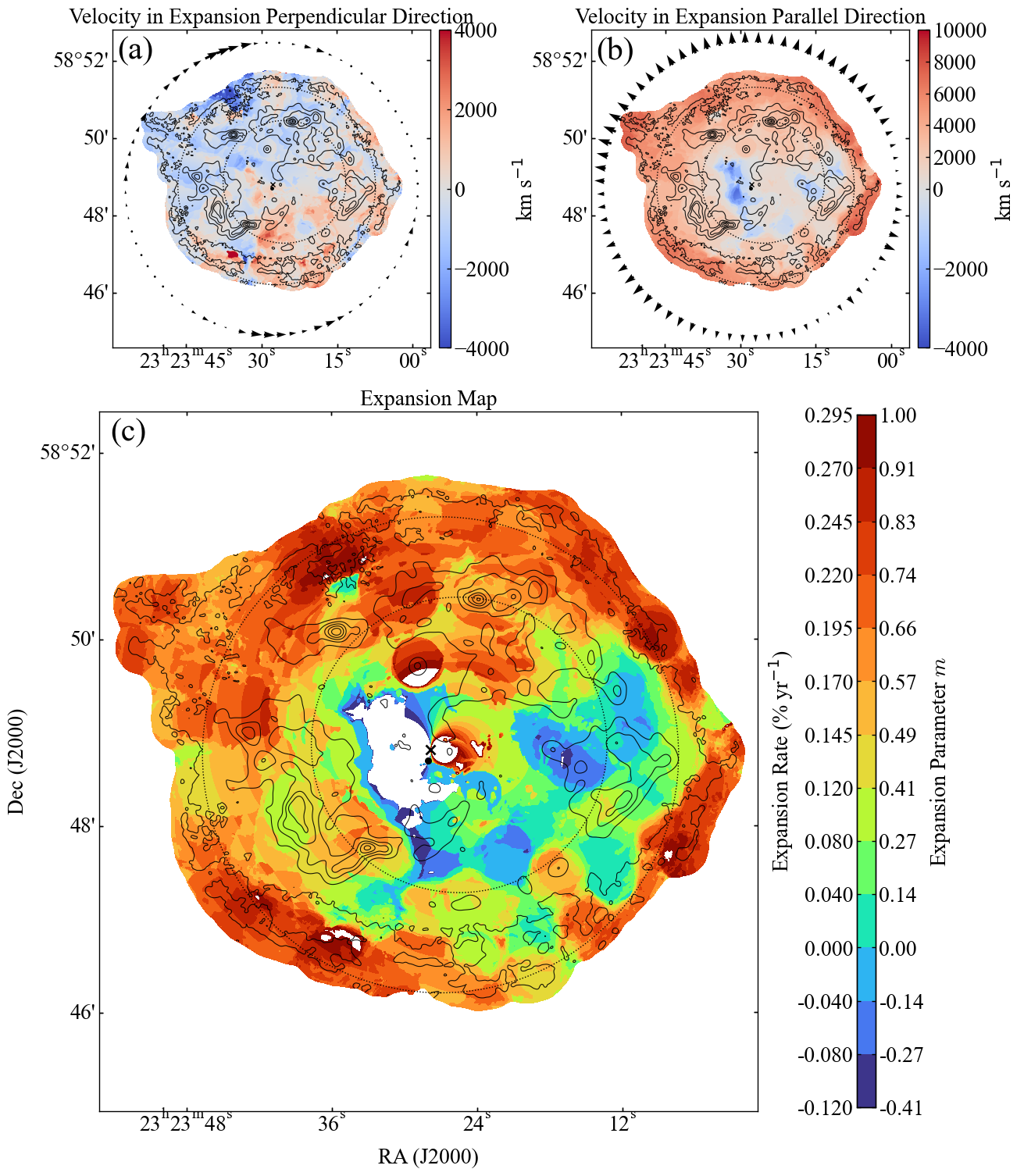}
 \caption{Velocity maps perpendicular and parallel to the expansion direction and expansion parameter. (a): Velocity perpendicular to the expansion direction, positive counterclockwise. (b): Velocity in the direction of expansion, positive outward. (c): Expansion parameter map calculated from (b) divided by the distance from the center of the explosion, i.e., the expansion rate. The expansion parameter is calculated assuming 338 years have passed since the explosion. White areas within Cas A indicate that the expansion rate is less than \SI{-0.12}{\%.yr^{-1}} or greater than \SI{0.295}{\%.yr^{-1}}. The vectors in (a) and (b) show the average vector intensity of the profiles for each 5$^\circ$ azimuthal angle from the explosion center. The contours in each image represent 2009, the CCO is indicated by a dot, the explosion center is marked by a cross, and the dotted circles indicate the forward and reverse shock positions.}
 \label{expansion_map}
\end{figure}

The proper motion in Figure~\ref{proper_motion}(b) is decomposed into two directions from the center of the explosion: the perpendicular direction in Figure~\ref{expansion_map}(a) with positive counterclockwise, and the parallel direction in Figure~\ref{expansion_map}(b) with positive outward.
For visual understanding of orientation, the average intensity for each 5$^\circ$ azimuthal angle is displayed as a vector. Note that the scaling intensities displayed in Figure~\ref{expansion_map}(a, b) are different.
In Figure~\ref{expansion_map}(a), some of the north-northeast and south-southwest forward shocks show a unique perpendicular motion of 1500--4000 \si{km.s^{-1}}. 
The south-southwest is dominated by dense CSM, as reported by \citet{Vink_2024}, so the shock wave could interact with this component. The north-northeast, with such perpendicular movements, may also have experienced some interaction or collision during the expansion.
For Figure~\ref{expansion_map}(b), the inward motion corresponds to almost the same region as identified by \citet{Sato_2018}.

The parallel velocity in Figure~\ref{expansion_map}(b) is used to calculate the expansion parameter by dividing it by the distance from the explosion center. Figure~\ref{expansion_map}(c) shows the expansion rate and the expansion parameter, which indicates deceleration from free expansion, assuming 338 years have passed in the 2009 data frame. Note that the regions deviating from the expected expansion rate, with an expansion rate of less than \SI{-0.12}{\%\ yr^{-1}} or greater than \SI{0.295}{\%\ yr^{-1}}, are highlighted in white. The results confirm the general trend of the expansion parameter $m\approx$ 0.66--0.74 (expansion rate of 0.195--0.220\si{\%\ yr^{-1}}), which is consistent with previous studies \citep[e.g.,][]{delaney2003first,patnaude2009proper,vink2022forward}. In the south and north shock regions, the $m\approx$ 0.57--0.66 (expansion rate of 0.170--0.195\si{\%\ yr^{-1}}), indicating that the deceleration is highly effective and consistent with previous findings \citep[e.g.,][]{delaney2003first,patnaude2009proper,vink2022forward}. 
These two velocity directions and the expansion parameter help us understand the kinematic properties (e.g., utilised in the parameters of the clustering method in Section~\ref{Classification of Kinematic Properties}).

\subsection{Relationship between Deceleration Component and CSM/ISM}\label{Relationship Between Deceleration Component and CSM/ISM}
\begin{figure}[ht!]
 \centering
 \includegraphics[width=1\linewidth]{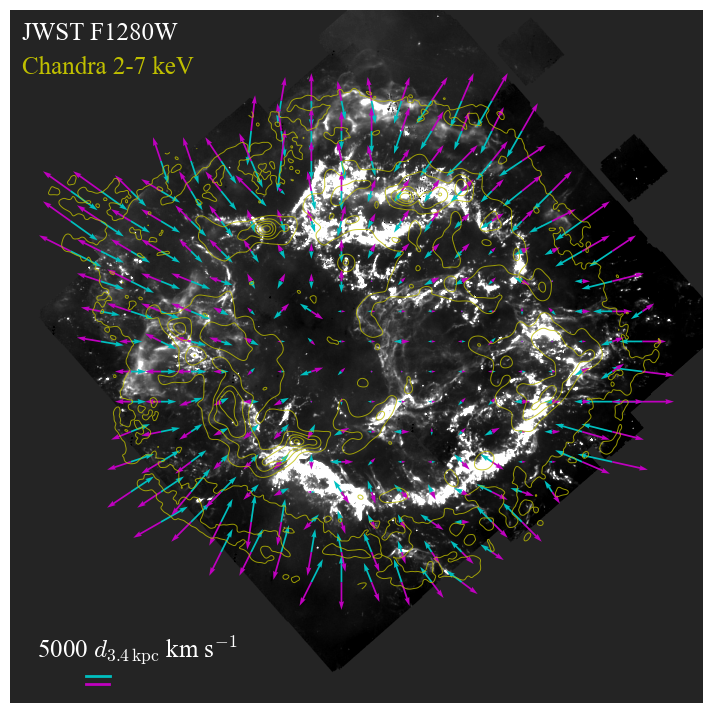}
 \caption{The Chandra proper motions superimposed on JWST observations. The grayscale image is from the F1280W filter of the JWST observation, while the yellow contours are based on Chandra 2--7 keV band image. The vectors represent the constant-velocity MLE-3 of Figure~\ref{proper_motion}(b), normalized to 50 year units and displayed at 40 pixel intervals.}
 \label{proper_motion_with_jwst}
\end{figure}

To investigate the interactions with the CSM/ISM, we compared the Chandra proper motions with JWST observations. The JWST data from 2022 August 4, were utilized in this study\footnote{The JWST data were obtained from \url{https://jwstfeed.com/}.}. Figure~\ref{proper_motion_with_jwst} illustrates the Chandra constant-velocity MLE-3 with the JWST F1280W filter overlaid. The contours are derived from Chandra data from 2009. The cyan and magenta vectors represent those shown in Figure~\ref{proper_motion}(b), with 40-pixel intervals on a 50-year scale. Focusing on the outer forward shock regions, a trend of qualitatively smaller Chandra proper motion is visible in the brighter areas of JWST.

Several characteristic areas in Figure~\ref{proper_motion_with_jwst} are discussed. In the outer southwest region, the proper motions are slower than the forward shocks, as reported by \citet{vink2022forward} and discussed in \citet{Vink_2024}, due to the higher CSM density in this region. The proper motions in Figure~\ref{proper_motion_with_jwst} demonstrate this trend and also indicate that the velocity direction tends to shift from the center to the counterclockwise direction when observing the perpendicular velocity (see details in Figure~\ref{expansion_map}(a)), likely due to the impact on the dense CSM. According to \citet{orlando2022evidence}, interactions with dense CSM can significantly affect the morphology and dynamics of SNRs, supporting our observations of deceleration and changes in velocity direction.

In the northern and northwestern regions \citep[regions of panels 1 and 2 of Figure 3 in][]{Milisavljevic_2024}, the results are slower than those of the nearby regions. \citet{Milisavljevic_2024} reported that these areas contain interstellar dust and CSM clouds. The proper motions using the more statistically rich year 2004 data and the preselection of the calculation MLE domain are presented in Appendix~\ref{Comparison of MLE Computation Domain Uniform and Preselected}, which similarly indicates slower velocities in such regions. By combining our MLE approach with JWST observations, we can better understand the deceleration and destruction as the forward shock wave passes through the CSM/ISM across the entire region. Looking ahead, the X-Ray Imaging and Spectroscopy Mission \citep[XRISM;][]{tashiro2022xrism}, launched in September 2023, will provide velocity information in the line-of-sight direction. Combining XRISM data will enhance our understanding of the kinematics in three-dimensional space.

\subsection{Technique for Capturing Flux Variation}\label{Technique for Capturing Flux Variation}
The MLE described in Section~\ref{Proper Motion Estimation} assumes Poisson statistics and a constant photon flux. In young SNRs, however, the thermal and nonthermal flux is universally altered, e.g., due to the amplified and turbulent magnetic field caused by diffusive shock acceleration \citep{axford1977acceleration,krymskii1977regular,bell1978acceleration,blandford1978particle,kamijima2020fast} and the decay of the thermal component by adiabatic expansion \citep{Sato_2017b}. This means that the model side ($\Lambda$) and the observed side ($K$ in the case of MLE-2, $K^{(0)}$ and $K^{(2)}$ in the case of MLE-3) in Equations~(\ref{mle_eq}, \ref{mle_3_eq}) will have outliers that cannot be explained by statistical errors. From a statistical approach, we propose a method to detect such regions of flux variation.

We use the likelihood ratio \citep[e.g.,][]{1979ApJ...228..939C} evaluated within the framework of interval estimation of the model $\Lambda$ and observations $K^{(0)}$ and $K^{(2)}$ in the constant-velocity MLE-3 method. The likelihood ratio is usually utilized to estimate the error of the proper motion calculated by the MLE \citep[e.g.,][]{sato2017freely,Tsuchioka_2021}, but we use it to evaluate the reliability of the model for Poisson statistics by considering the degrees of freedom for each pixel of the $\Lambda$.
When the elements of the region $\Omega$ for the model $\Lambda$ are free parameters, the likelihood ratio is represented as
\begin{equation}
\begin{split}
\chi^2 &\sim -2\ln \frac{\mathcal{L}_{\textrm{MLE-3}}\left(\hat{\bm{\Delta}}^{(10)}_{x,y},\,\hat{\bm{\Delta}}^{(12)}_{x,y};\, \Lambda,\, K^{(0)},\, K^{(2)}\right)}{\mathcal{L}_{\textrm{MLE-3}}\left(\hat{\bm{\Delta}}^{(10)}_{x,y},\,\hat{\bm{\Delta}}^{(12)}_{x,y};\, \hat{\Lambda},\, K^{(0)},\, K^{(2)}\right)}\\
&=-2\ln\frac{\mathcal{L}\left(\hat{\bm{\Delta}}^{(10)}_{x,y};\, \Lambda^{(10)},\, K^{(0)}\right) \mathcal{L}\left(\hat{\bm{\Delta}}^{(12)}_{x,y};\, \Lambda^{(12)},\, K^{(2)}\right)}{\mathcal{L}\left(\hat{\bm{\Delta}}^{(10)}_{x,y};\, \hat{\Lambda}^{(10)},\, K^{(0)}\right) \mathcal{L}\left(\hat{\bm{\Delta}}^{(12)}_{x,y};\, \hat{\Lambda}^{(12)},\, K^{(2)}\right)},
\label{chi_eq}
\end{split}
\end{equation}
where $\hat{\bm{\Delta}}^{(10)}_{x,y}$ and $\hat{\bm{\Delta}}^{(12)}_{x,y}$ are the maximum likelihood solutions of the $\mathcal{L}_{\textrm{MLE-3}}(\cdot)$ in Equation~\eqref{mle_3_eq}. Under the determination of that solution, $\hat{\Lambda}$ represents the parameters that maximize the $\mathcal{L}_{\textrm{MLE-3}}(\cdot)$ function. Since $\Lambda$ has common parameters for $K^{(0)}$ and $K^{(2)}$ via $N^{(1)}$ and $A^{(1)}$ in the epoch-1 image, we introduce the commonized parameters $\hat{N}^{(1)}$ and $\hat{A}^{(1)}$. From the property of the Poisson distribution, the maximum value is obtained as the mean of the pixels shifted by $\hat{\bm{\Delta}}^{(10)}_{x,y}$ and $\hat{\bm{\Delta}}^{(12)}_{x,y}$, which is described as
\begin{equation}
\begin{split}
\hat{N}^{(1)}_{x,y}&\coloneqq N^{(0)}_{x+\hat{\Delta}^{(10)}_x,y+\hat{\Delta}^{(10)}_y}+N^{(2)}_{x+\hat{\Delta}^{(12)}_x,y+\hat{\Delta}^{(12)}_y}\\
\hat{A}^{(1)}_{x,y}&\coloneqq A^{(0)}_{x+\hat{\Delta}^{(10)}_x,y+\hat{\Delta}^{(10)}_y}+A^{(2)}_{x+\hat{\Delta}^{(12)}_x,y+\hat{\Delta}^{(12)}_y}\\
\hat{\Lambda}^{(10)}_{x,y}&=\frac{{A}^{(0)}_{x+\hat{\Delta}^{(10)}_x,y+\hat{\Delta}^{(10)}_y}}{\hat{A}^{(1)}_{x,y}}\hat{N}^{(1)}_{x,y}\\
\hat{\Lambda}^{(12)}_{x,y}&=\frac{{A}^{(2)}_{x+\hat{\Delta}^{(12)}_x,y+\hat{\Delta}^{(12)}_y}}{\hat{A}^{(1)}_{x,y}}\hat{N}^{(1)}_{x,y}.
\label{chi_eq_param}
\end{split}
\end{equation}
This equation can be treated as $\chi^2$ \citep{1979ApJ...228..939C}, where the degrees of freedom are the number of elements in $\Omega$.

\begin{figure}[ht!]
 \centering
 \includegraphics[width=1\linewidth]{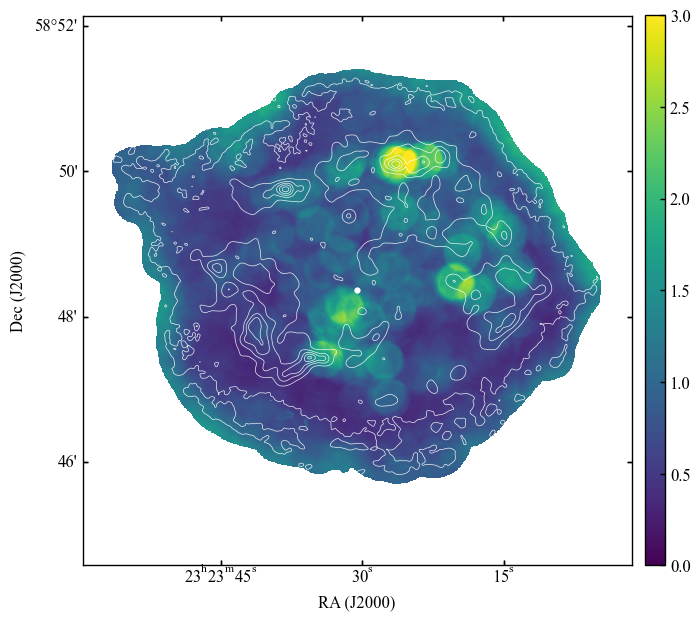}
 \caption{Likelihood ratio map computed with the constant-velocity MLE-3. This map is divided by the degrees of freedom and represents a reduced $\chi^2$. The degrees of freedom are the size of the computational domain of the MLE method, which is 65 pixels in diameter, or 3209.}
 \label{likelihood_ratio}
\end{figure}

Figure~\ref{likelihood_ratio} illustrates the likelihood ratio map of Equation~\eqref{chi_eq} divided by the degrees of freedom, representing a reduced $\chi^2$. The degrees of freedom here is 3209, corresponding to the number of elements in the computational domain $\Omega$ with a diameter of 65 pixels. The result demonstrates the goodness of fit to the model, where a value close to 1 indicates statistical correctness. This identification allows us to pinpoint areas that cannot be adequately explained by the Poisson model.

We discuss the physical interpretation of regions with high likelihood ratios in Figure~\ref{likelihood_ratio}. First, we compare with nonthermal components and knots. They are consistent with rapid acceleration and cooling by amplified magnetic fields therein \citep[e.g.,][]{uchiyama2008fast,patnaude2009proper}. They also correspond to brightness variations such as knots and small-scale clumps \citep[e.g.,][]{patnaude2006small,patnaude2014comparison}, and contain complex motion components caused by inward reverse shocks with significant geometric changes \citep[e.g.,][]{Sato_2018}.

In contrast, the variation in thermal flux to the southeast reported by \citep{Sato_2017b} is not evident in our results. This difference arises because the average statistics in our analysis assume 10--30 photons per pixel, resulting in a statistical error of approximately 18--32\%. Consequently, detecting variations within this range is statistically challenging. Given that the typical synchrotron radiation flux varies by approximately \SI{4}{\%.yr^{-1}} \citep{uchiyama2008fast} and thermal radiation by about \SI{1}{\%.yr^{-1}} \citep{Sato_2017b}, detecting variations in thermal radiation over a decade is quite difficult. However, we have observed fluctuations in thermal radiation in the northern reverse shocks \citep{Sato_2017b}. This is because the northern region exhibits a complex three-dimensional velocity component \citep{delaney2010three}, which could capture intricate geometrical variations that may not be replicated in two-dimensional models. Additionally, the northern reverse shocks are believed to experience stronger deceleration due to CSM interaction \citep{Wu_2024}, contributing to changes in the brightness and/or spatial distribution of the flux. Therefore, this method, although reliant on statistical and geometrical conditions, has the potential to identify previously unknown regions in a data-driven manner.

\subsection{Optimal Computational Domain Size Estimation}\label{Optimal Computational Domain Size Estimation}
The size of the calculation domain $\Omega$ in Section~\ref{Proper Motion Estimation} is a crucial hyper-parameter for the stability of the MLE method. A smaller domain allows for the detection of local motion but increases the influence of Poisson noise in sparse regions. Conversely, a larger domain facilitates the tracking of global motions but presents challenges in capturing localized structures. This section describes a method for assessing the accuracy of motion estimation using a global objective measure. 
The method is based on the relative deviation between the source image and the image shifted by the computed proper motion, determined by finding the minimum point. Additionally, we will confirm that the RL$_{\mathrm{sv}}$ method improves accuracy over observations.

The MLE method computes proper motion vectors for each pixel in the image over the computational domain, resulting in a complete set of vectors corresponding to the image size.
To assess the overall correctness of the MLE across a comprehensive area of Cas~A, we compare the differences between the source image and the image shifted according to proper motion, as suggested by \citet{ichinohe2023spatiotemporal}.
Here, the shifted images are derived from the source image, with each pixel's motion corresponding to the results obtained in Section~\ref{Result of the MLE Method}. 
To measure and compare the difference between the original image "No move" and the transformation "Move" by proper motion of two side 2009 to 2000 and 2009 to 2019, we define the relative deviation as
\begin{equation}
\begin{split}
\textrm{No move: }&\frac{1}{2}\left(\frac{|F_{2009}-F_{2000}|}{F_{2000}} +\frac{|F_{2009}-F_{2019}|}{F_{2019}}\right)\\
\textrm{Move: }&\frac{1}{2}\left(\frac{|F_{2009\to2000}-F_{2000}|}{F_{2000}} +\frac{|F_{2009\to2019}-F_{2019}|}{F_{2019}}\right),
\end{split}
\label{relative_deviation_eq}
\end{equation}
where shifted images from 2009 to 2000 and from 2009 to 2019 are denoted as $F_{2009\to2000}$ and $F_{2009\to2019}$, respectively. The original images from 2000, 2009, and 2019, without any motion processing, are labeled as $F_{2000}$, $F_{2009}$, and $F_{2019}$.

\begin{figure}[ht!]
 \centering
 \includegraphics[width=1\linewidth]{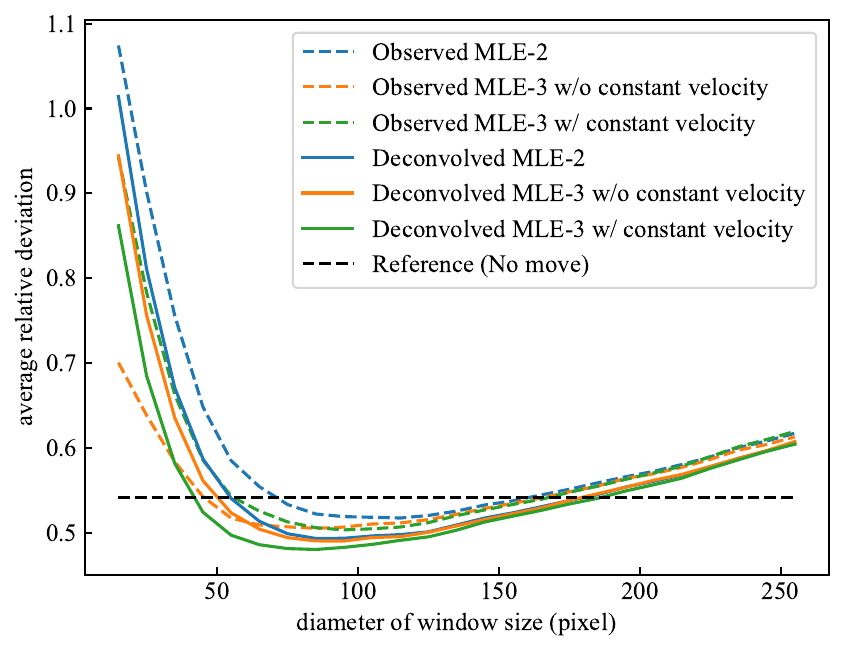}
 \caption{Relationship between the relative deviation and the computational domain size of MLEs. Blue line: the mean value of Equation~(\ref{relative_deviation_eq}: Move) using the MLE-2 movement. Orange and green lines: same as the blue line, except for the MLE-3 without and with constant velocity movements, respectively. The dashed and solid lines show the MLE method using the observed image with 6 pixel sigma Gaussian smoothing and RL$_{\mathrm{sv}}$-deconvolved with 1.5 pixel sigma Gaussian smoothing, respectively. The black dashed line represents the average of Equation~(\ref{relative_deviation_eq}: No move) as the reference line.}
 \label{windowsize_estimate}
\end{figure}

Figure~\ref{windowsize_estimate} shows the relationship between the relative deviation and the computational domain size in the moved image created by the proper motion of the MLE method. The calculation size is displayed every 10 pixels, from 15 to 255 pixels. The dashed and solid lines in Figure~\ref{windowsize_estimate} show the MLE method using the observed image with 6 pixel sigma Gaussian smoothing and the smoothed-RL$_{\mathrm{sv}}$ image, respectively. The reason for the smoothing in the observed image is to account for the effect of the PSF caused by the different optical axis positions of each observation (the choice of the Gaussian sigma parameter is based on the off-axis PSF; see Appendix~\ref{Appropriate smoothed Gaussian sigma of RLsv}). 
In order to align and evaluate the observation and deconvolution of MLEs, the images used for $F_{2000}$, $F_{2009}$, and $F_{2019}$, as well as the images moved to create $F_{2009\to2000}$ and $F_{2009\to2019}$ in Equation~\eqref{relative_deviation_eq}, are the smoothed-RL $_{\mathrm{sv}}$ images.

The blue, orange, and green lines in Figure~\ref{windowsize_estimate} represent the average of Equation~(\ref{relative_deviation_eq}: Move) using the motion of MLE-2 and MLE-3 without and with constant velocity, respectively. It can be seen that MLE-3 shows smaller values than MLE-2 in both observations and deconvolution, indicating higher accuracy. For small computational domains, the curves are more complex owing to the strong influence of factors such as smoothing effects and PSF differences. The convergence positions show that the observed values are roughly between 85 and 95 pixels, while the deconvolution values converge to a smaller range of about 75 to 85 pixels. Additionally, the minimum coefficient of variation for the deconvolution results is the smallest compared to the others, indicating the highest accuracy. The black dashed line represents the average of Equation~(\ref{relative_deviation_eq}: No move) as a reference line; values falling below this line indicate relatively adequate movement. All types of MLEs fall below the baseline at the minimum point. The size of the computational domain with the smallest relative deviation is around 75 pixels for deconvolution, and since empirical results around this value have been good, this can be considered a way to determine the optimal scale of the calculation domain.

\begin{figure*}[ht!]
 \centering
 \includegraphics[width=1\linewidth]{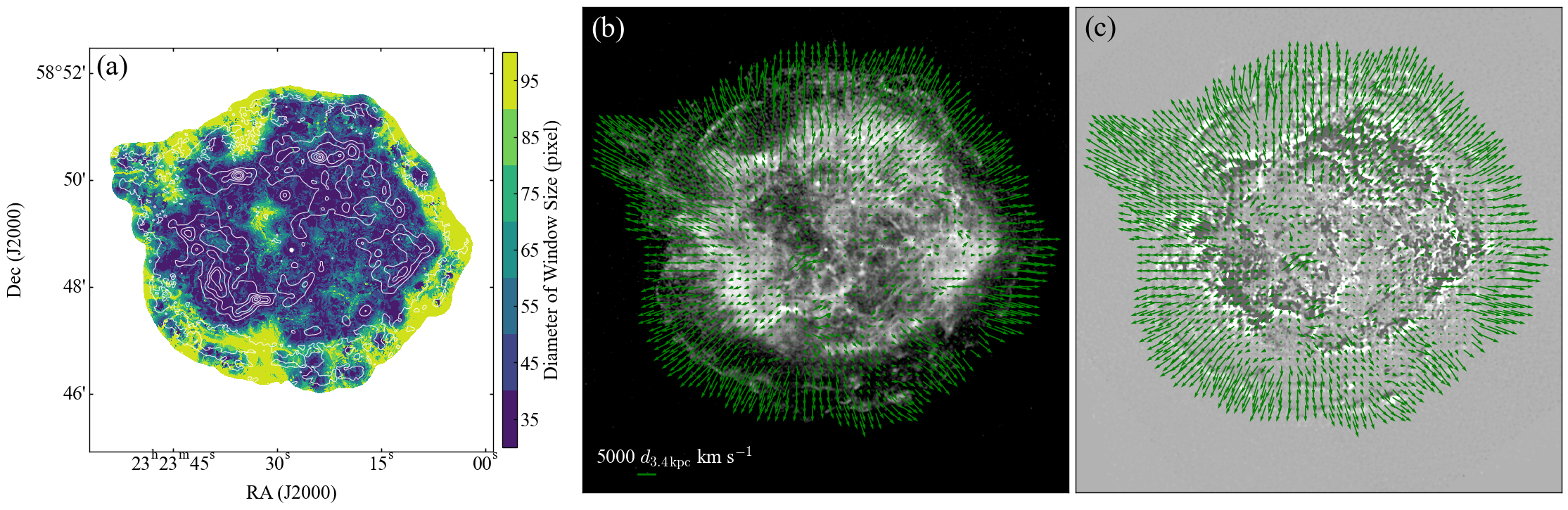}
 \caption{Results of the constant-velocity MLE-3 method with multiple computation domains. (a): Diameter size of the computational domain selected based on the standard deviation of the proper motion. The contours correspond to the year 2009, and the CCO is marked with a dot. (b): Vector field calculated from the domain size in (a). These vectors show the 2009--2019 side, normalized to 50-year units and displayed at 15-pixel intervals. The background image is from 2009. (c): Same vectors as in (b) overlaid on the difference images for 2000 and 2019. (An animation of this figure is available in ApJ. The animation consists of three frames, with only the superimposed images for the years 2000, 2009, and 2019 switched in Figure~\ref{multi_proper_motion}(b). The similarity between the estimated proper motion vector and the actual image motion can be confirmed.)}
 \label{multi_proper_motion}
\end{figure*}

\subsection{Further Improvement Approach}\label{Further Improvement Approach}
In this section, we propose two techniques aimed at enhancing the constant-velocity MLE-3 method for improved extraction and comprehension of the kinematic features of Cas~A. Section~\ref{MLE with Appropriate Computational Domains} discusses the combination of multiple computational domains to improve sensitivity to motion at different scales. Section~\ref{Classification of Kinematic Properties} introduces a classification of kinematic properties to examine the characteristics of Cas A using various features such as proper motion vectors and expansion rates, as detailed in Section~\ref{Verification of MLE Results with Previous Studies}.

\subsubsection{MLE with Appropriate Computational Domains}\label{MLE with Appropriate Computational Domains}
Section~\ref{Optimal Computational Domain Size Estimation} described a method for determining the parameters of the computational domain $\Omega$. However, in Cas~A, with its complex motion components, the spatial scale of motion should differ from place to place. Additionally, even in cases of localized motion, the aperture problem \citep[e.g.,][]{shimojo1989occlusion,barron1994performance} needs to be addressed, especially in homogeneous regions. Therefore, a possible approach to capture such movements with high sensitivity is to use an appropriately scaled $\Omega$ for each location.

We propose a straightforward implementation of MLE-3 with position-dependent computational domains. Our strategy for multi-domain MLE-3 is to prioritize the proper motions obtained with high confidence in smaller computational domains. This approach means that small computational domains will result in large errors when Poisson noise and aperture problems occur. In such areas, a larger domain should be employed. Conversely, motion measured robustly in a small computational domain is assumed to capture true local features and should be actively utilized. The method employs circular regions with diameters of 35, 45, 55, 65, 75, 85, and 95 pixels, which are determined to include the optimal size in the uniform case (see details in Section~\ref{Optimal Computational Domain Size Estimation}). The reason for combining them at 10-pixel intervals is to reduce computational cost: at 35 pixels, finer motion can be estimated, and as the number increases to 95 pixels, global features can be captured. The appropriate combination of these domains has the potential to measure any type of motion.

The procedure for combining computation domains is described below. First, calculate the standard deviation of the constant-velocity MLE-3 for each computational domain (see Appendix~\ref{Procedure for Calculating the Standard Deviation of Velocity} for details on the calculation of the standard deviation). Next, set the threshold at 0.5 pixels of standard deviation, or 1-sigma per pixel. Starting at 35 pixels, obtain proper motion below the threshold value. Then, for the areas not captured at 35 pixels, proper motion within the threshold value is used at 45 pixels. This process is repeated up to 85 pixels. Finally, 95 pixels are used for all remaining regions.

The result of applying this selection rule is shown in Figure~\ref{multi_proper_motion}. Figure~\ref{multi_proper_motion}(a) illustrates the diameter sizes of the computational domains adopted through this process. Bright areas are segmented into 35 or 45-pixel domains, while dark areas and structural edges are assigned domains of 55 to 85 pixels. Regions between structures are marked with a domain of 95 pixels. Figure~\ref{multi_proper_motion}(b) displays the proper motions for the 2009--2019 side using the computational domains shown in Figure~\ref{multi_proper_motion}(a). The vectors are normalized to represent a 50-year period and are displayed at 15-pixel intervals. The background image in Figure~\ref{multi_proper_motion}(b) is the 2009 image. The difference image between 2000 and 2019 is shown in Figure~\ref{multi_proper_motion}(c) to qualitatively validate the motion. This multi-domain MLE tends to detect local structures along the difference image (Figure~\ref{multi_proper_motion}(c)) and clearly shows improved sensitivity compared to using the uniform domain case in Figure~\ref{proper_motion}(b).

\begin{figure*}[ht!]
 \centering
 \includegraphics[width=1\linewidth]{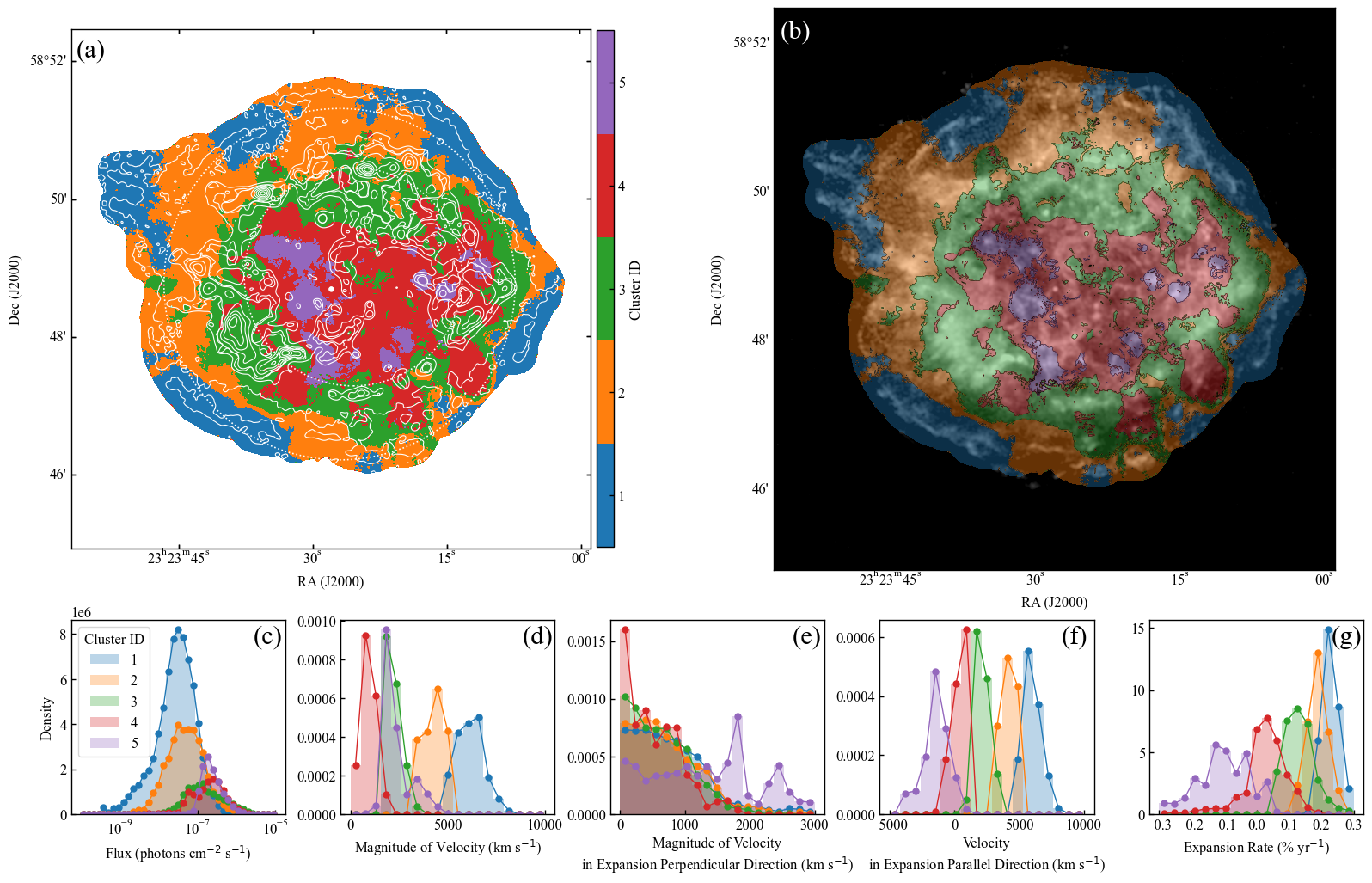}
 \caption{Classification results with 5 clusters using the $k$-means algorithm. The input parameters are the smoothed-RL$_{\mathrm{sv}}$ image of 2009, magnitude of velocity, magnitude of velocity in the perpendicular direction from the explosion center, velocity in the parallel direction from the explosion center, and expansion rate. The parameters (except for the flux parameter) are calculated from the proper motion result of Figure~\ref{multi_proper_motion}. (a): Each cluster is indicated by its own color: Cluster ID=1 (blue), 2 (orange), 3 (green), 4 (red), and 5 (purple), with 2009 contours overlaid. The CCO is marked with a dot, and two dotted circles show the positions of forward and reverse shock waves. (b): Same as (a), but using the 2009 image as the background. (c--g): Histogram of input parameters calculated for each classified cluster. To improve visibility, the bins of the histogram are connected by lines. (An animation of this figure is available in ApJ. The animation consists of three frames, with only the overlaid images for the years 2000, 2009, and 2019 switched in Figure~\ref{clustering_kmeans}(b). Visual confirmation of the consistency of movement for each clustering can be observed.)}
 \label{clustering_kmeans}
\end{figure*}
\begin{table*}[ht!]
\caption{Classification Results of $k$-means Algorithm}
\centering
\setlength{\tabcolsep}{2.5pt} %This is to prevent the table from sticking out, but change it accordingly.
\begin{tabular}{cccccc}
\hline
Cluster ID & Flux & Velocity Magnitude & Velocity Magnitude & Velocity & Expansion Rate \\
 &  &  &   Perpendicular Direction &  Parallel Direction & \\
& (\SI{e-7}{photons.cm^{-2}.pixel^{-2}.s^{-1}}) & (\si{km.s^{-1}})& (\si{km.s^{-1}}) & (\si{km.s^{-1}}) & (\si{\%.yr^{-1}}) \\ \hline
1 & $1.17\pm 1.23$ & $6138\pm 708$ & $860\pm 768$ & $6031\pm 687$ & $0.228\pm 0.026$\\
2 & $3.65\pm 5.06$ & $4280\pm 547$ & $730\pm 534$ & $4182\pm 561$ & $0.188\pm 0.033$\\
3 & $8.79\pm 9.05$ & $2242\pm 489$ & $639\pm 470$ & $2089\pm 524$ & $0.136\pm 0.059$\\
4 & $5.77\pm 4.84$ & $880\pm 414$ & $518\pm 422$ & $368\pm 603$ & $0.028\pm 0.095$\\
5 & $5.06\pm 4.76$ & $2374\pm 840$ & $1477\pm 981$ & $-1457\pm 1038$ & $-0.238\pm 0.257$\\
\hline
\end{tabular}
\label{clustering_kmeans_table}
\end{table*}

\subsubsection{Classification of Kinematic Properties}\label{Classification of Kinematic Properties}
We introduce the $k$-means algorithm to understand the kinematic characteristics of Cas~A.
The $k$-means method uses the following five input parameters: the flux value of the smoothed-RL$_{\mathrm{sv}}$ image, the magnitude of the proper motion velocity, the magnitude of the perpendicular velocity measured from the explosion center, the parallel velocity from the explosion center, and the expansion rate (A simple classification of the $k$-means with only position and proper motion as parameters is described in Appendix~\ref{Spatial Relationship of Proper Motion}). The reason for employing the absolute value rather than the velocity itself for the perpendicular direction is to parameterize motion along or away from the expansion direction. The first input parameter is from the 2009 image, while the others are from the 2009--2019 proper motion results obtained in Section~\ref{MLE with Appropriate Computational Domains}. The calculation procedure for the last three parameters related to the explosion center is described in Section~\ref{Comparison with SNR Evolution Model}. These parameters are removed from the outer masked region indicated in Section~\ref{Result of the MLE Method}, min-max normalized, and entered into the $k$-means method. The number of clusters is set to 5, considering empirical results with different parameters and physical interpretability.

The results of applying the $k$-means method are shown in Figure~\ref{clustering_kmeans}. The Cluster IDs 1, 2, 3, 4, and 5 correspond to the colors blue, orange, green, red, and purple, respectively. Figures~\ref{clustering_kmeans}(a, b) represent the obtained clusters superimposed on the contours and image of the smoothed-RL$_{\mathrm{sv}}$, respectively. Figures~\ref{clustering_kmeans}(c--g) show the histograms of each cluster after transforming the input parameters to their pre-normalized state. To improve visibility, each bin is connected by lines. The means and standard deviations of the parameters for these clusters are listed in Table~\ref{clustering_kmeans_table}.

We identify the characteristics of these five clusters. First, as an overall feature, the Cluster IDs are aligned from 1 to 5, from the outside to the inside. Focusing on each cluster individually, Cluster ID=1 correlates well with the relatively high-velocity ($\gtrapprox$6000 \si{km.s^{-1}}; see Figure~\ref{clustering_kmeans}(d)) forward shocks measured in \citet{vink2022forward}. It also shows the northeast jet and the southwest counter-jet \citep{hwang2004million,vink2022forward,ikeda2022discovery}. Furthermore, it seems to capture the empty zones or less clumpy structures of the northeast-southwest pair \citep[Funnel 1 and Funnel 2 in][]{bear2024identifying}. Therefore, we identify Cluster ID=1 as fast forward shocks and jet-like structures with relatively little interaction with the CSM.

Focusing mainly on the northern and southern regions, Cluster ID=2 is found to be rich in interstellar dust and CSM clouds observed by JWST \citep{Milisavljevic_2024, Vink_2024}. With a velocity magnitude of \SI{\approx 4000}{km.s^{-1}} and an expansion rate of \SI{\approx 0.19}{\%.yr^{-1}} (indicating that the expansion parameter $m < 2/3$), we conclude that Cluster ID=2 captures forward shocks with high CSM interaction.

Cluster ID=3 predominantly occupies the reverse shock regions \citep{gotthelf2001chandra}, as indicated by the inner dotted circle in Figure~\ref{clustering_kmeans}(a). As shown in Table~\ref{clustering_kmeans_table}, the velocity map confirms its outward motion at about \SI{2200}{km.s^{-1}}, with an expansion rate of \SI{\approx 0.136}{\%.yr^{-1}}. These velocity and expansion rates are broadly consistent with those reported for the outgoing reverse shock by \citet{vink2022forward}. We therefore classify Cluster ID=3 as the outgoing reverse shock region.

Cluster ID=4 has the lowest velocity among these clusters at $\approx 800$ $\si{km.s^{-1}}$ (see Table~\ref{clustering_kmeans_table}). This velocity closely matches the typical velocity range of quasi-stationary flocculi \citep[QSFs;][]{van1971optical, koo2020detection}, which are typically $\lessapprox 550$ \si{km.s^{-1}} \citep{delaney2004kinematics,koo2020detection}. In fact, the cluster spans the southwestern region of QSFs \citep{koo2020detection}. Additionally, it includes the very slow inward/outward western reverse shock regions. This association suggests that Cluster ID=4 is related to the QSFs component and the much slower reverse shock region.

The perpendicular velocity in Figure~\ref{clustering_kmeans}(e) for Cluster ID=5 indicates movement away from the explosion center. Additionally, Figure~\ref{clustering_kmeans}(f) illustrates its motion toward the center. Notably, this center and the western region are closely aligned with the region described by \citet{Sato_2018}. It also captures the westward movement to the south near the CCO and the movement toward the center in the northeastern region close to the CCO, as noted by \citet{patnaude2009proper}. Furthermore, a complex pattern of northwestward motion is observed on the northeastern side near the CCO within this cluster (see Figure~\ref{multi_proper_motion}(c) for the difference image and proper motions). This suggests that Cluster ID=5 represents the reverse shock moving inward and a complex motion that cannot be explained by simple expansion.

By fusing MLE-3 and $k$-means methods, kinematic information can be classified into physical components and mapped onto the image.
Although the choice of clustering algorithms and input parameters cannot be addressed within the scope of this paper, it is expected to develop a classification of kinematic features that includes, e.g., the use of proper motion for other energy bands, and/or multiwavelength images.

\section{Conclusion}\label{Conclusion}
We implemented the multiepoch MLE method with three time-series images (MLE-3 method) and demonstrated its effectiveness on the SNR Cas~A using Chandra data from 2000, 2009, and 2019. The performance improvement by using three observations and applying the RL$_{\mathrm{sv}}$  method was quantitatively confirmed by images shifted based on the MLE results. We also quantitatively evaluated the credibility of the velocity field by comparing it to previous studies. Additionally, the velocity was decomposed into parallel and perpendicular components to investigate motion deviating from the expansion, revealing regions where motion clearly deviates from the expansion and/or moves inward. By overlaying our proper motion result on the JWST observations, we qualitatively investigated the deceleration of the forward shock wave in the CSM/ISM dominant region.

Furthermore, we advanced this method by combining multiple computational regions and conducting unsupervised classification using parameters such as proper motion, photon flux, and expansion rate. This approach allowed us to classify the physical components, such as interactions with the CSM and the inward-moving reverse shock, and to visualize the asymmetric structure in a data-driven way. These techniques enable us to observe the dynamics of SNRs from diverse perspectives and uncover their unique evolution.

This research has made use of data obtained from the Chandra Data Archive and the Chandra Source Catalog and software provided by the Chandra X-ray Center (CXC) in the application packages CIAO. 
This work was supported by the JSPS KAKENHI grant No. 24KJ2067.

%\cleardoublepage
\appendix

\section{Position Dependence of PSF}\label{Position dependence of PSF}
\begin{figure*}[ht!]
 \centering
 \includegraphics[width=1\linewidth]{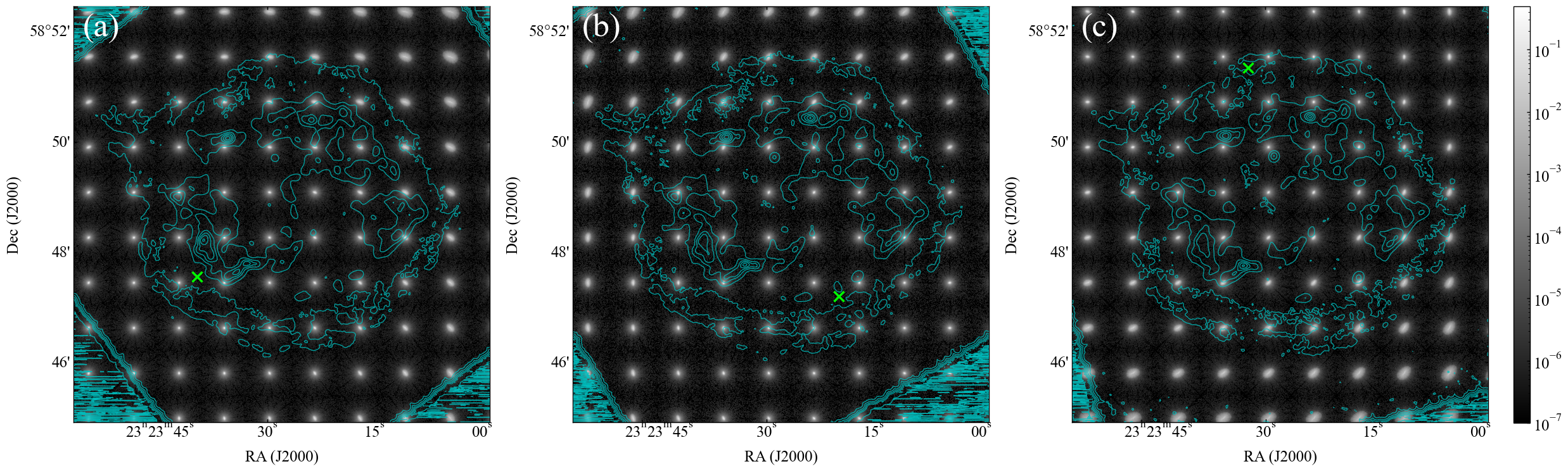}
 \caption{PSFs for each location are arranged at 100$\times$100 pixel intervals. (a), (b), and (c) represent Obs.~ID=114, 10935, and 19606, respectively. The integral of each PSF is normalized to 1, and a uniform color range is applied. The position of the optical axis is marked with a cross. The contours for each panel in (a, b, c) are images of the years 2000, 2009, and 2019, respectively.}
 \label{psfs_map}
\end{figure*}

To illustrate the characteristics of Chandra's PSF, we present the results with the corresponding positions and PSFs. Figures~\ref{psfs_map}(a, b, c) show the PSFs generated every 100$\times$100 pixels by \texttt{simulate\_psf} for the 2000, 2009, and 2019 observations, respectively, overlaid at their locations. Note that the RL$_{\mathrm{sv}}$ in this paper used a PSF of 25$\times$25 pixels, but for clarity the PSF is shown here at these intervals. The optical axis is marked with a cross, indicating that it is different for each year. The PSF shows minimum diffusion at the optical axis. As the position deviates from the optical axis, the PSF tail expands, accompanied by a gradual shift of the elliptical axis.

\section{Appropriate Smoothed Gaussian Sigma after RL$_{sv}$ Method}\label{Appropriate smoothed Gaussian sigma of RLsv}
The RL$_{\mathrm{sv}}$ method introduces systematic errors that vary due to PSF variations at different locations. This uncertainty becomes more pronounced off-axis. For example, if a point source is observed both on-axis and off-axis, applying the RL method will likely result in the former converging more accurately to a point, thus better representing the true sky image. Naturally, realistic observations involve various components beyond point sources and necessitate more complex considerations. In this study, we propose a straightforward approach to estimate and mitigate this effect using the RL method with the PSF itself.

\begin{figure}[ht!]
 \centering
 \includegraphics[width=1\linewidth]{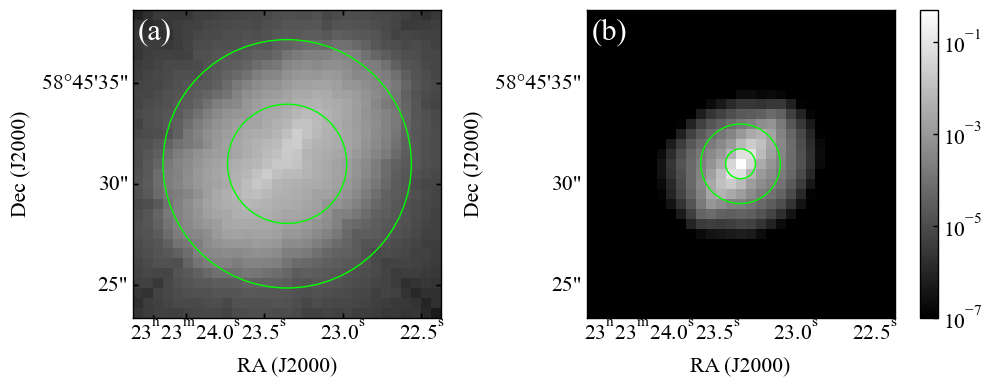}
 \caption{(a): The off-axis PSF at monochromatic energy 3.8 keV in Obs.~ID=19606. The integral of the PSF is normalized to be 1. (b): Result of 50 iterations of the RL method with the same PSF as in (a). The color scale is shared, and the inner and outer circles represent 1-sigma and 2-sigma confidence intervals, respectively.}
 \label{off_axis_psf_self_rl}
\end{figure}

We use the off-axis (southwest region) PSF for 2019 (Obs. ID=19606) simulated by \texttt{simulate\_psf}. Figure~\ref{off_axis_psf_self_rl}(a) illustrates this PSF, centered in the image. Figure~\ref{off_axis_psf_self_rl}(b) displays the results of RL method using this PSF as the observed image. The two circles in each panel of Figure~\ref{off_axis_psf_self_rl} represent the probability densities of 1-sigma and 2-sigma: (a) 1-sigma of 6 pixels and 2-sigma of 12.5 pixels; (b) 1-sigma of 1.5 pixels and 2-sigma of 4 pixels. This result aids in estimating the systematic error that does not return to the correct position. In this paper, to reduce the off-axis uncertainty in the RL$_{\mathrm{sv}}$ method, we smooth the RL$_{\mathrm{sv}}$ images (Figures~\ref{rlsv}(d, e, f)) using a Gaussian function with a sigma of 1.5 pixels based on 1-sigma in Figure~\ref{off_axis_psf_self_rl}(b).

\section{Procedure for Calculating the Standard Deviation of Velocity}\label{Procedure for Calculating the Standard Deviation of Velocity}
We present a procedure for evaluating fluctuations in the constant-velocity MLE-3 method caused by statistical uncertainties. The evaluation process of the statistical fluctuations is as follows: First, Poisson noise is added to each of the RL$_{\mathrm{sv}}$ images from 2000, 2009, and 2019. Next, the constant-velocity MLE-3 method is applied to these generated images after smoothing with a Gaussian sigma of 1.5 pixels. This smoothing accounts for both the off-axis PSF uncertainty and the reduction of Poisson noise. Finally, the procedure is repeated 100 times to obtain the standard deviation of the proper motions. The standard deviations are defined as the scalar quantity $\sqrt{\sigma_x^2 + \sigma_y^2}$, where $\sigma_x$ and $\sigma_y$ are the standard deviations in arbitrary $(x,\,y)$ coordinates of 100 motion sets, computed individually for the $x$- and $y$-directions. The standard deviation is calculated using the 2009--2019 side because the values are almost identical to those in 2009--2000 when using the constant-velocity MLE-3 framework.

\section{Comparison of MLE Computation Domain Uniform and Preselected}\label{Comparison of MLE Computation Domain Uniform and Preselected}
We examine the estimation accuracy by comparing the MLE method in a uniform computational domain (as shown in Section~\ref{Result of the MLE Method}) with that in a preselected computational domain. 
In the preselection case, the MLE-2 method is used as a simplified framework to minimize systematic errors, and the statistically rich 2004 data is adopted to reduce statistical errors.

\begin{figure}[ht!]
 \centering
 \includegraphics[width=1\linewidth]{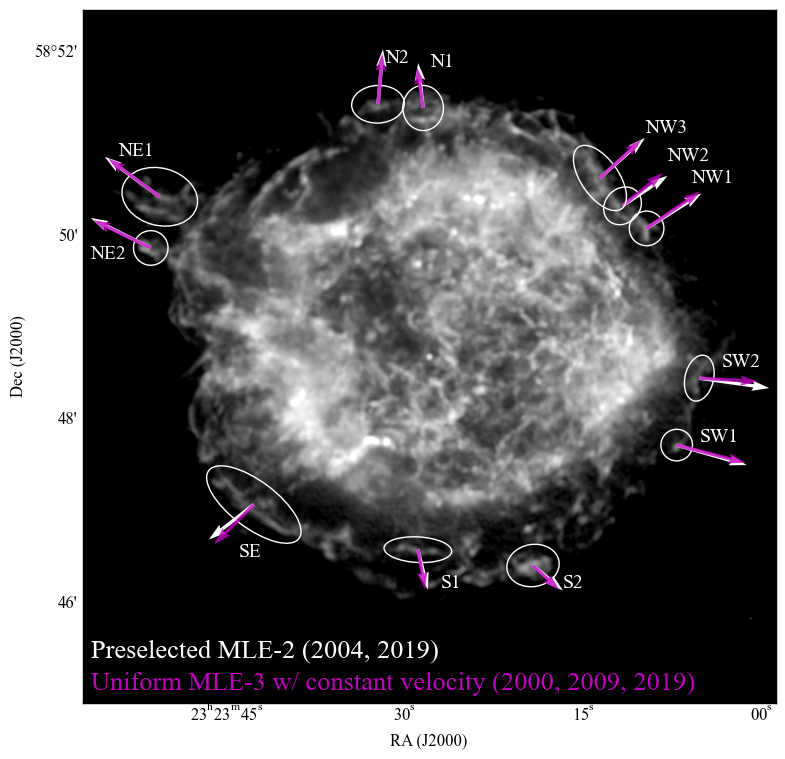}
 \caption{Comparison of the results of the uniform and preselected computation domains of MLE. The ellipse in the image represents the preselected computational domain. The proper motion vectors from 2009 to 2019, averaged over the elliptical region for the 2000, 2009, and 2019 constant-velocity MLE-3 with uniform calculation domain (following the same procedure as in Figure~\ref{proper_motion}(b)), are shown as magenta vectors. For the preselected case, the MLE-2 method uses statistically rich data from 2004 and data from 2019 in the ellipse domains, with the results shown by white vectors. All vectors are scaled to 100 years. The background image represents the smoothed RL$_{\mathrm{sv}}$ image from 2004.}
 \label{proper_motion_uniform_vs_preselected_fig}
\end{figure}

Figure~\ref{proper_motion_uniform_vs_preselected_fig} illustrates the preselected regions visually identified (indicated by the white oval). For the uniform computational domain MLE, the constant-velocity MLE-3 from 2000, 2009, and 2019, as described in Section~\ref{Result of the MLE Method}, is used. The motion vector from 2009 to 2019 is obtained for each pixel, and the average value within the white area is calculated. These mean vectors are displayed in magenta. For the preselected case, the proper motions for each preselected domain are calculated by MLE-2 ($\Lambda=2004$ and $K=2019$ in Equation~\eqref{mle_eq}). Each proper motion vector is scaled to a 100-year, and the background image shows the smoothed RL$_{\mathrm{sv}}$ image from 2004. Although strictly speaking, the MLE results for the 2009 reference should be adjusted to align the region with the 2004 reference; for simplicity, this adjustment is not applied to the calculation regions.

\begin{table}[ht!]
\caption{Proper Motion Results for MLE Method with a Uniform Domain vs. Preselected Domain MLE}
\centering
\begin{tabular}{ccc}
\hline
Region$^a$ & Uniform Domain$^b$ & Preselected Domain$^c$ \\
& (\si{km.s^{-1}})& (\si{km.s^{-1}}) \\ \hline
N1 & 4325$\pm$465 & 4767$\pm$372\\
N2 & 5341$\pm$834 & 5815$\pm$372\\
NE1 & 6758$\pm$699 & 7160$\pm$372\\
NE2 & 6708$\pm$779 & 7063$\pm$372\\
SE & 5887$\pm$1257 & 6002$\pm$372\\
S1 & 4250$\pm$450 & 4341$\pm$372\\
S2 & 3795$\pm$589 & 4112$\pm$372\\
SW1 & 7481$\pm$656 & 7665$\pm$372\\
SW2 & 6158$\pm$1200 & 7445$\pm$372\\
NW1 & 6886$\pm$498 & 6864$\pm$372\\
NW2 & 5510$\pm$812 & 5694$\pm$372\\
NW3 & 5902$\pm$786 & 6339$\pm$372\\
\hline
\end{tabular}
\footnotesize{\\$^a$: The regions are indicated in Figure~\ref{proper_motion_uniform_vs_preselected_fig}. \\$^b$: The magnitude of the velocity and its error are the mean and standard deviation of the proper motions in Figure~\ref{proper_motion}(b) (constant-velocity MLE-3 in 2000, 2009, and 2019). \\$^c$: The magnitude of the velocity is calculated by the MLE-2 method (2004 and 2009) in the selected region, and the error is computed by the $C$-statistic.}
\label{proper_motion_uniform_vs_preselected_table}
\end{table}

Table~\ref{proper_motion_uniform_vs_preselected_table} provides the velocity magnitudes for each region in Figure~\ref{proper_motion_uniform_vs_preselected_fig}. The uniform region shows the mean velocity and its standard deviation, while the preselected region presents the proper motions and the corresponding 1-sigma values calculated using the $C$-statistic \citep{1979ApJ...228..939C}. The $C$-statistic results indicate that the 1-sigma confidence intervals fall within $\pm 0.5$ pixels in all regions, and considering the two directions, the Euclidean distance $\pm 0.5\sqrt{2}$ pixels is used as the error. The table indicates that the measured velocities in each region agree within the margin of error.
One notable area is N1 and NW2, which are rich in CSM/ISM \citep{Milisavljevic_2024} and are measurably slower than the neighboring regions N2, NW1, and NW3 by both approaches. 

\section{Spatial Relationship of Proper Motion}\label{Spatial Relationship of Proper Motion}
We introduce the $k$-means algorithm to understand the characteristics of proper motion at each location. The method uses velocity fields of multi-domain MLE-3 from the 2009--2019 side obtained in Section~\ref{MLE with Appropriate Computational Domains}. The input parameters for $k$-means are the proper motion vector $(\Delta_x,\, \Delta_y)$ and its position $(x,\, y)$ for each pixel. These parameters are removed from the outer background mask region specified in Section~\ref{Result of the MLE Method}. They are then pre-processed by min-max normalization for the motions $(\Delta_x,\, \Delta_y)$ and positions $(x,\, y)$, respectively, and fed into the $k$-means method. The number of clusters for $k$-means is chosen so that each cluster approximately covers a circular area of about 75 pixels in diameter. The diameter size is the optimal computational domain for this MLE, as explained in Section~\ref{Optimal Computational Domain Size Estimation}. Consequently, the number of clusters is set to 100. 

\begin{figure}[ht!]
 \centering
 \includegraphics[width=1\linewidth]{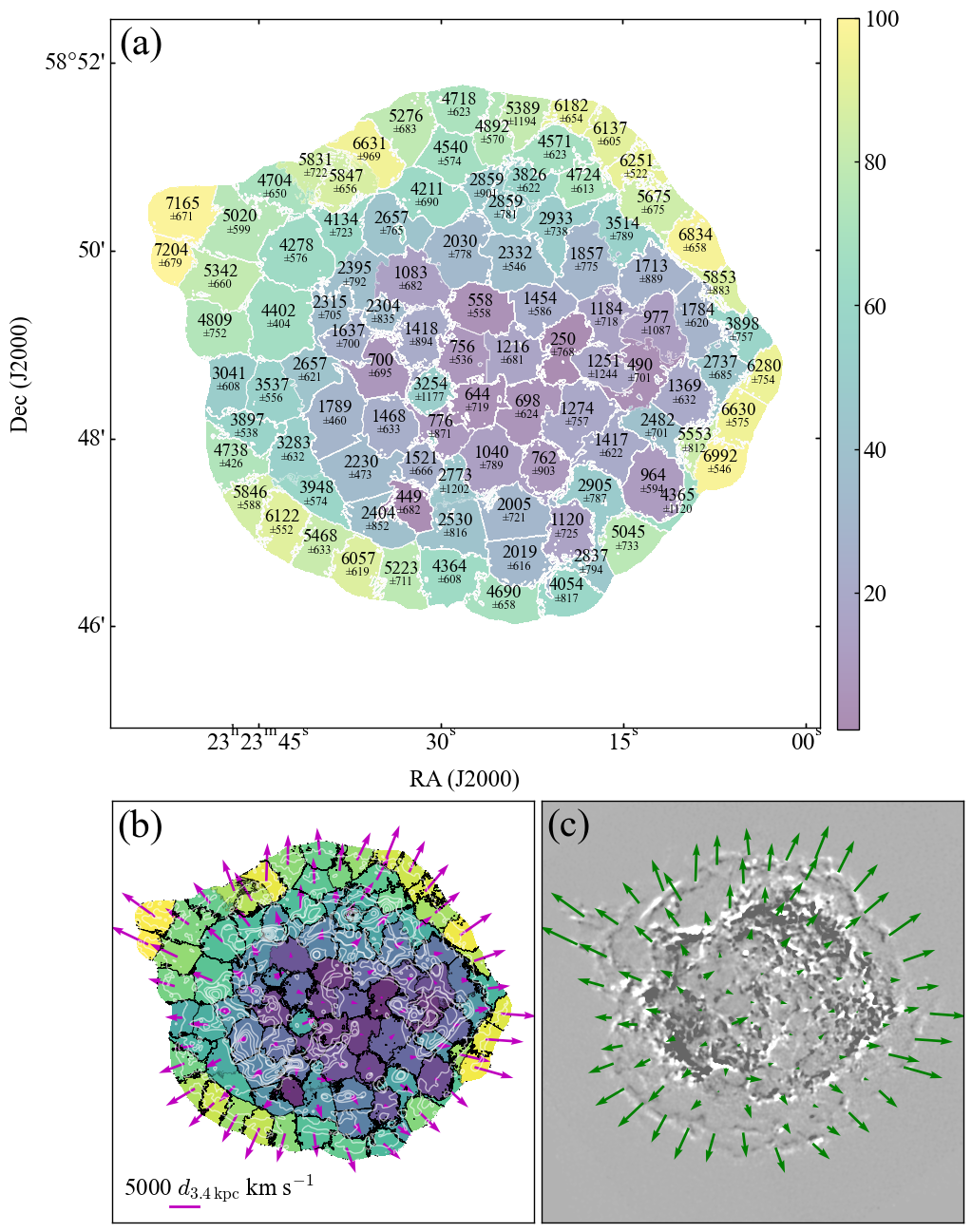}
 \caption{Classify the results of constant-velocity MLE-3 with multiple computation domains (Figure~\ref{multi_proper_motion}) using the $k$-means method. (a): Showing the $k$-means with 100 clusters sorted by velocity magnitude and displayed on a color scale with the mean velocity and its error. (b): The mean velocity vector in each cluster on a 100-year scale, with 2009 contours overlaid. (c): The same vectors as in (b), with the difference between 2000 and 2019 superimposed.}
 \label{clustering_cell}
\end{figure}

Figure~\ref{clustering_cell}(a) illustrates the $k$-means result, sorted by velocity magnitude. At each cluster's centroid, the velocity magnitude and its error are displayed, determined by calculating the mean and standard deviation of the velocity components in the x- and y-dimensions using the Euclidean distance. Figure~\ref{clustering_cell}(b) shows the same clusters as in Figure~\ref{clustering_cell}(a), with their mean proper motions and the 2009 contours overlaid. Figure~\ref{clustering_cell}(c) presents the same vectors as in Figure~\ref{clustering_cell}(b), superimposed with the difference images from 2000 and 2019. From Figure~\ref{clustering_cell}, spatial motion characteristics can be observed from both global and local perspectives. Adjacent clusters of similar color indicate robust kinematic associations, while abrupt color transitions between clusters indicate a complex amalgamation of motion components or structural boundaries.

\bibliography{MLE_apj}{}
\bibliographystyle{aasjournal}

\end{document}